\documentclass[CEJP,PDF]{cej}
\usepackage{layout}
\usepackage{amsmath}
\usepackage{textcomp}
\usepackage{hyperref}

\newcommand{\be}{\begin{equation}}
\newcommand{\ee}{\end{equation}}
\newcommand{\bea}{\begin{eqnarray}}
\newcommand{\eea}{\end{eqnarray}}
\newcommand{\eq}{\begin{equation}}
\newcommand{\eqx}{\end{equation}}
\newcommand{\eqn}{\begin{eqnarray}}
\newcommand{\bi}{\begin{itemize}}
\newcommand{\eqnx}{\end{eqnarray}}
\newcommand{\ei}{\end{itemize}}
\newcounter{hran}

\newcommand{\ba}{\begin{array}}
\newcommand{\ea}{\end{array}}
\newcommand{\balg}{\begin{align}}
\newcommand{\ealg}{\end{align}}

\newcommand{\lsim}
{\raise0.3ex\hbox{$\;<$\kern-0.75em\raise-1.1ex\hbox{$\sim\;$}}}
\newcommand{\gsim}
{\raise0.3ex\hbox{$\;>$\kern-0.75em\raise-1.1ex\hbox{$\sim\;$}}}

\title{Beyond the Standard Model of Physics with Astronomical Observations}

\articletype{Research Article}

\author{Raul Jimenez\inst{1}$^,$\inst{2}\email{raul.jimenez@icc.ub.edu}}

\institute{
\inst{1} ICREA \& ICC, University of Barcelona (IEEC-UB), Marti i Franques 1, Barcelona 08028, Spain
\inst{2} Theory Group, Physics Department, CERN, CH-1211, Geneva 23, Switzerland}

\abstract{There has been significant recent progress in observational cosmology. This, in turn, has provided an unprecedented picture of the early universe and its evolution.  In this review I will present a (biased) view of how one can use these observational results to constraint fundamental physics and in particular physics beyond the standard model.}

\begin{document} 

\maketitle

\section{Introduction}

Over the last two decades we have seen phenomenal progress in our observational understanding of the Universe. It was in the early 90s that the first  image of the fluctuation of the Cosmic Microwave Background (CMB) was obtained by the COBE satellite. It was around the same time that the first survey of galaxies was carried out by the CfA. Twenty years later we have had our knowledge of the cosmos revolutionised by the outstanding data collected by two galaxy surveys: 2dF and SDSS and by two CMB satellites WMAP \& Planck, which have provided both full sky temperature and polarization measurements of the CMB. In fact, Planck has already obtained all information in the temperature power spectrum that is available in the scale (cosmic variance limited) at angular scales larger than 5'. While these datasets have had a profound impact on our understanding of the  cosmos, they also provide an invaluable tool to extract information about fundamental physics. The future is even brighter, because there are current and future cosmological experiments planned to map out the whole sky at all wavelengths. Further, the promise of detecting gravitational waves should be realised during this decade with advanced ground-based observatories like LIGO, and therefore a new window will open at mapping the Universe. In this talk, I will give a (presenter biased) view of some insights into fundamental physics that have been obtained using cosmological observations. Namely: constraints on dark energy, neutrino masses and hierarchy and beyond the standard model physics. Most of the material shown here has been presented elsewhere in many of my own referred publications, but is presented here coherently to give a panchromatic view of how to shed light on beyond the standard model physics with cosmology. 

\section{Dark energy}

\begin{table}[b!]
\begin{center}
\begin{tabular}{||c||ccc|cc||ccc|cc||}
\hline
& \multicolumn{5}{c||}{BC03 models} & \multicolumn{5}{|c||}{MaStro models} \\
$z$ & $H(z)$ & $\sigma_{stat}$ & $\sigma_{syst}$ & $\sigma_{tot}$ & \% error & $H(z)$ & $\sigma_{stat}$ & $\sigma_{syst}$ & $\sigma_{tot}$ & \% error\\
\hline
0.1791 & 75 & 3.8 & 0.5 & 4 & $5\%$ & 81 & 4.1 & 2.5 & 5 & $6\%$\\
0.1993 & 75 & 4.9 & 0.6 & 5 & $7\%$ & 81 & 5.2 & 2.6 & 6 & $7\%$\\
0.3519 & 83 & 13 & 4.8 & 14 & $17\%$ & 88 & 13.9 & 7.9 & 16 & $18\%$\\\
0.5929 & 104 & 11.6 & 4.5 & 13 & $12\%$ & 110 & 12.3 & 7.5 & 15 & $13\%$\\
0.6797 & 92 & 6.4 & 4.3 & 8 & $9\%$ & 98 & 6.8 & 7.1 & 10 & $11\%$\\
0.7812 & 105 & 9.4 & 6.1 & 12 & $12\%$ & 88 & 8 & 7.4 & 11 & $13\%$\\
0.8754 & 125 & 15.3 & 6 & 17 & $13\%$ & 124 & 14.3 & 8.7 & 17 & $14\%$ \\
1.037 & 154 & 13.6 & 14.9 & 20 & $13\%$ & 113 & 10.1 & 11.7 & 15 & $14\%$\\
\hline
\end{tabular}
\caption{$H(z)$ measurements (in units of [$\mathrm{km\,s^{-1}Mpc^{-1}}$]) and their errors; the columns in the middle report
the relative contribution of statistical and systematic errors, and the last ones the total error (estimated by summing in quadrature $\sigma_{stat}$ and $\sigma_{syst}$). These values have been estimated respectively with BC03 and MaStro stellar population synthesis models. This dataset can be downloaded at the address 
http://www.astronomia.unibo.it/Astronomia/Ricerca/ Progetti+e+attivita/cosmic\_chronometers.htm (alternatively http://start.at/cosmicchronometers).}
\label{tab:HzBC03}
\end{center}
\end{table}

 Direct supernova measurements of the deceleration parameter
\cite{Perlmutter:1998np}, as well as indirect
measurements based upon a combination of results from the cosmic
microwave background (CMB)
\cite{cmb}, large-scale structure
(LSS) \cite{percival2df,dodelson}, and the Hubble constant \cite{H0}
indicate that the  expansion is accelerating.  This suggests
either that gravity on the largest scales is described by some
theory other than general relativity and/or that the Universe is filled
with some sort of negative-pressure ``dark energy'' that drives
the accelerated expansion \cite{reviews}; either way, it requires new physics
beyond general relativity and the standard model of particle
physics.  These observations have garnered
considerable theoretical attention as well as
observational and experimental efforts to learn more about the new
physics coming into play.

\begin{figure}
\begin{center}
\includegraphics[width=\columnwidth]{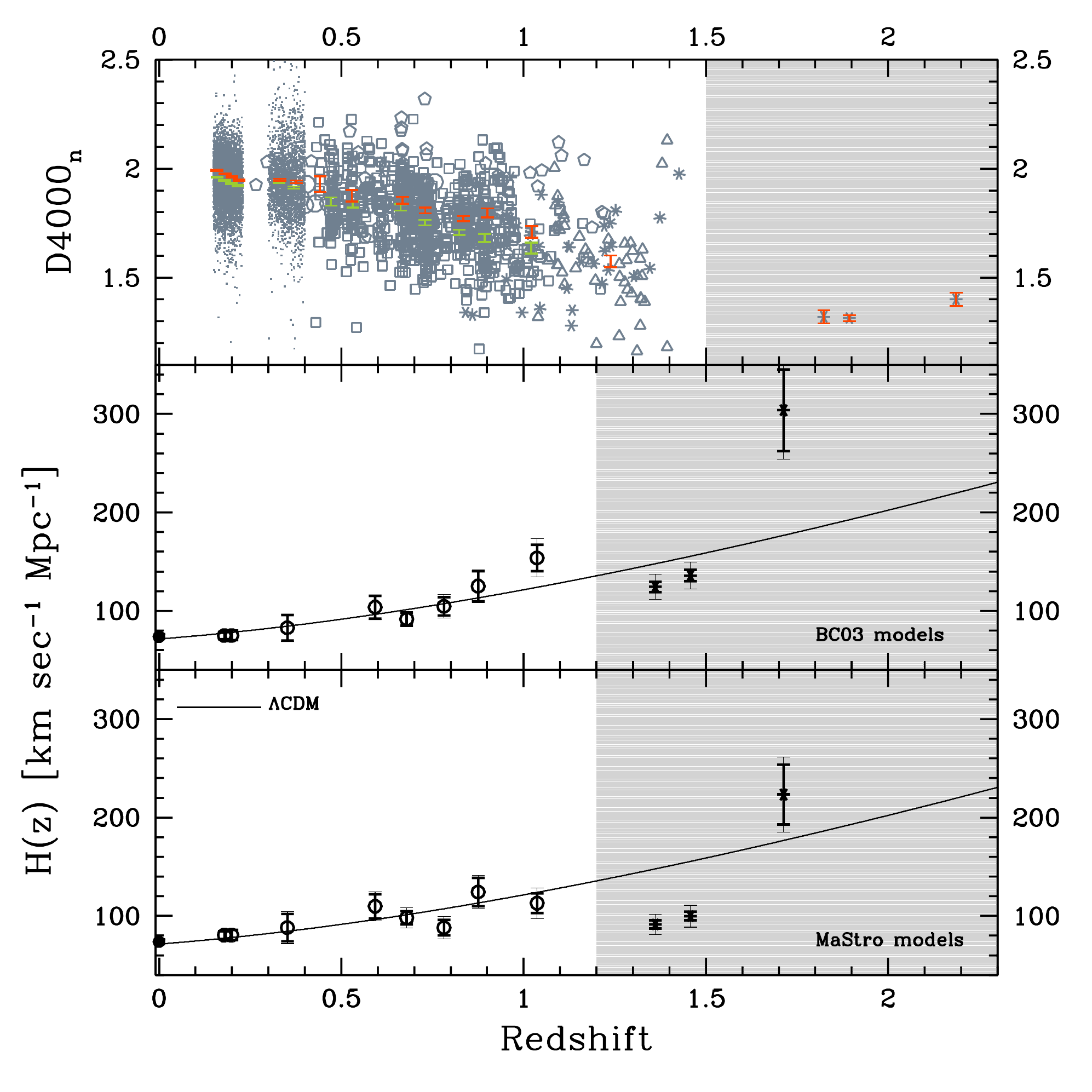}
\caption{\label{fig:Hz} The $H(z)$ measurement from a sample
     of passively evolving galaxies. The points at $z > 1$ are
     taken from Ref.~\cite{moresco}, while the point at $z=0$ is
     from the Hubble Key Project \cite{H0}. The solid line is the $\Lambda$CDM
     model.}
\end{center}
\end{figure}

\begin{figure}
\begin{center}
\includegraphics[width=\columnwidth]{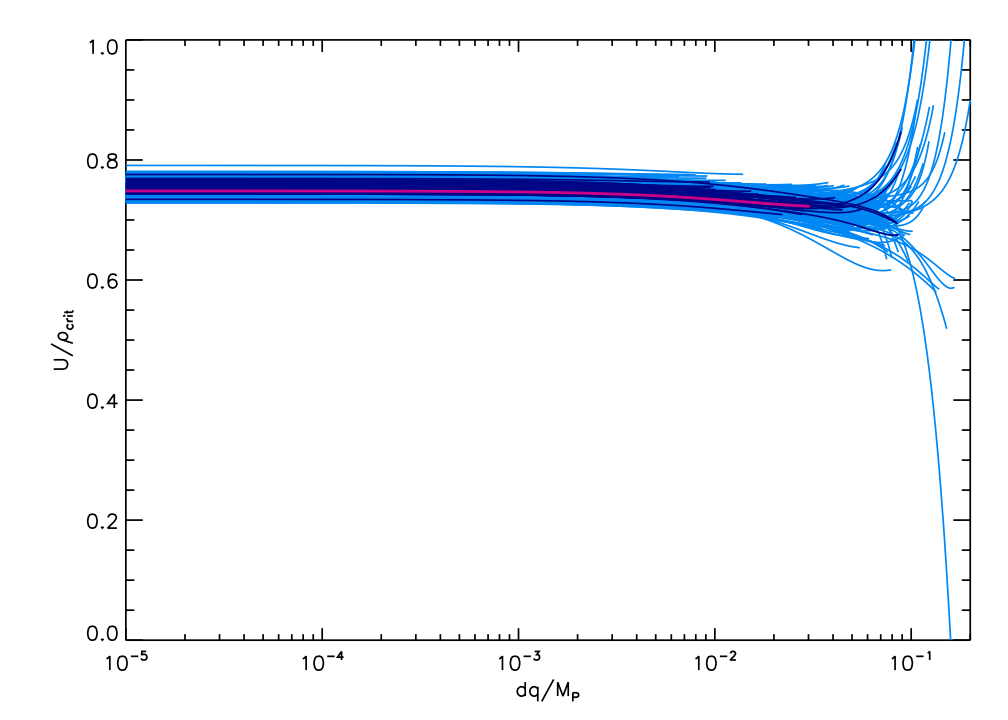}
\caption{\label{fig:depot} The effective potential of accelerated expansion $U(q)$ in units of the critical density $\rho_{\rm crit}$ as function of the displacement 
of the field $q$ in $M_{\rm p}$ units. Different tracks are plotted for values with 68\% confidence (dark blue lines) and 95\%  confidence (light blue lines). The best-fit model is shown as a solid red line, which is better described physically as a pseudo-Goldstone boson. The trajectories correspond to how much the field has moved in the full redshift range of the observational data ($0 < z < 1.1$).  Note that the potential is very flat at the few $6$\% level and that for many models the field displacement is very small. The data strongly favor a flat potential.}
\end{center}
\end{figure}

The simplest possibility is to extend Einstein's equation with a
cosmological constant, or equivalently, to hypothesize a fluid with
an equation-of-state parameter $w\equiv p/\rho = -1$ (with $p$ and
$\rho$ the pressure and energy density, respectively).  However,
it may well be that the cosmological ``constant'' actually evolves
with time, in which case $w\neq -1$, and there are a variety of
theoretical reasons \cite{ratrapeebles} to believe that this is the
case.  Precise measurement of $w(z)$ (with, in general, a parameterized
redshift dependence) or, equivalently, the cosmic expansion history,
has thus become a central goal of physical cosmology \cite{Peacock06,detf}.

Among the techniques to determine the cosmic expansion
history are supernova searches, baryon acoustic oscillations (BAO)
\cite{percival,eisenstein,BAO,pritchard}, weak lensing \cite{weaklensing}, and
galaxy clusters \cite{clusters}.  These techniques all have different strengths,
and they all also suffer from a different set of weaknesses.  As
argued in the ESO/ESA and Dark Energy Task Force reports \cite{Peacock06,detf}, robust
conclusions about the cosmic expansion history will likely
require several avenues to allow for cross checks.  There may
also still be room for other ideas for determining the expansion
history.

A common weakness of supernova searches, BAO (at least the
angular clustering), weak lensing, and cluster-based
measurements is that they rely largely on integrated
quantities.  For example  supernovae probe the luminosity
distance,
\begin{equation}
     d_L(z) =(1+z) \int_z^0 (1+z') \frac{dt}{dz'}dz'.
\label{eqn:dL}
\end{equation}
The other probes rely on similar quantities, which depend on an
integral of the expansion history, to determine the expansion
history, rather than the expansion history itself.  The purpose
of the differential-age technique \cite{jimenezloeb} is to
circumvent this limitation by measuring directly the integrand,
$dt/dz$, or in other words, the change in the age of the
Universe as a function of redshift.  This can be achieved 
by measuring ages of galaxies with respect to a fiducial 
model, thus circumventing the need 
to compute absolute ages. From Galactic globular clusters age-dating we 
know that relative ages are much more accurately determined than
absolute ages (e.g.,
Refs.~\cite{chaboyer,jimenez95,chaboyerkrauss03}).   A
preliminary analysis, with archival data, has already been
carried out \cite{jimenezstern,svj}, and the results applied to
constrain dark-energy theories
(e.g., Refs.~\cite{Hz1}).

The challenge with the differential-age measurement is to find a population of
standard clocks and accurately date them. There is now growing
observational evidence that the most massive galaxies contain
the oldest stellar populations up to redshifts of $z \sim 1-2$
\cite{dunlop96,spinrad,Cowie99,Heavens04,thomas}. Refs.~\cite{Heavens04} and
\cite{panter} have shown that the most massive galaxies have
less than 1\% of their present stellar mass formed at $z <
1$. Ref.~\cite{thomas} shows that star formation in massive systems in
high-density regions --- i.e., galaxy clusters --- ceased by redshift
$z \sim 3$ and Ref.~\cite{treu} shows that massive systems, those with
stellar masses $ > 5 \times 10^{11}$~M$_{\odot}$, have finished
their star-formation activity by $z \sim 2$.
There is thus considerable empirical evidence for a
population of galaxies, harboured in the highest-density regions
of galaxy clusters, that has formed its stellar population at
high redshift, $z > 2$, and that since that time this population has
been evolving passively, without further episodes of star
formation. These galaxies trace the
``red envelope,'' and are the oldest objects in the Universe
at every redshift.  The differential ages of these
galaxies should thus be a good indicator for the rate of change
of the age of the Universe as a function of redshift.

The most recent measurements of the expansion history
obtained from the ages of passively-evolving galaxies in galaxy
clusters  at $z<2.0$ have been reported by \cite{sternjim,moresco,morescoII} (see Fig.~1 and Table~1).  Such
observations provide a promising new cosmological constraint,
particularly for understanding the evolution of the dark-energy
density (Fig.~\ref{fig:depot}). The current measurements already provide valuable constraints \cite{morescoIII,JTVMCP,MJCP},
and the success of this effort should motivate further
measurements along these lines as well as a more intensive
investigation of the theoretical underpinnings of the
calculations and the associated uncertainties.  Further, there
has been significant advancement in the last few years in modelling
stellar populations of LRG galaxies and
the differential technique has been recently applied
very successfully to determine the metallicity of LRGs.

{\em The main highlight  is that  the expansion history of the Universe is consistent  with that predicted by a flat potential, i.e. a cosmological constant. The data do not require extra parameters  beyond a constant term in the Lagrangian to explain the current accelerated expansion}. Further, deviations in the potential from a constant are constrained to be below $6$\%.  Observational constraints allow the parameters describing the Lagrangian to vary only within certain limits; the  relative range of  the allowed variation of the  parameters confirms a  well defined hierarchy where the linear and quadratic terms dominate over higher-order terms, justifying the  basic assumption of the effective theory approach. Observational constraints also give some indications of the relevant energy scales involved.   Because a direct determination of a Lagrangian allows  us to determine the underlying symmetries in the theory, our results can be used to shed light on this as well. 

Additional effort on both the theoretical and observational
side  may ultimately promote the differential-age technique
as an important new dark-energy avenue which complements supernova
searches, weak lensing, baryon acoustic oscillations, and
cluster abundances. The differential-age technique can potentially provide {\em model independent} measurements of the expansion history of the universe at the \% level \cite{Bellini,Putter,HTJHCM,HJM}. Also, absolute ages of stellar ages have proven very valuable to establish possible deviations from the LCDM model \cite{local1,local2}

\section{Measuring the neutrino mass and its hierarchy}

In the past decade, there has been great progress in neutrino
physics.  It has been shown that 
atmospheric neutrinos exhibit a large up-down asymmetry in the SuperKamiokande (SK) 
experiment. This was the first significant evidence for a finite neutrino mass \cite{SuperK} 
and hence the incompleteness of the Standard Model of particle physics.
Accelerator experiments \cite{K2K, MINOS} have confirmed this evidence and improved the 
determination of the neutrino mass splitting required to explain the observations.
The Sudbury Neutrino Observatory (SNO) experiment 
has shown that the solar neutrinos change their flavors from the electron type to other active 
types (muon and tau neutrinos)\cite{SNO}. Finally, the KamLAND reactor anti-neutrino oscillation
experiments reported a significant deficit in reactor anti-neutrino flux over approximately 180~km 
of propagation \cite{KamLAND}. Combining results from the pioneering
Homestake experiment \cite{Homestake} and
Gallium-based experiments \cite{Gallium}, the decades-long solar
neutrino problem \cite{solarproblem} has been solved by   the electron
neutrinos produced at Sun's core propagating adiabatically to a heavier
mass eigenstate due to the matter effect \cite{MSW}.

As a summary, two hierarchical neutrino mass splittings and two large mixing angles have been 
measured, while only a bound on a third mixing angle has been established.
Furthermore the standard model has three neutrinos and the motivation 
for considering deviations from the standard model in the form of 
extra neutrino species has now  disappeared \cite{mena,miniboone}.

New neutrino experiments aim to determine the remaining parameters of the neutrino mass 
matrix and the nature of the neutrino mass. Meanwhile, relic neutrinos  produced in the early universe are 
hardly detectable by weak interactions but new cosmological probes offer the opportunity to detect 
relic neutrinos and determine neutrino mass parameters.

It is very relevant that the maximal mixing of the solar mixing angle is excluded at a 
high significance. The exclusion of the maximal mixing by SNO \cite{SNO} has an important impact on a deep  question in neutrino physics: ``are neutrinos their own
anti-particle?".  If the answer is yes, then neutrinos are Majorana fermions; if not,
they are Dirac. If neutrinos and anti-neutrinos are identical, there could have been a
process in the Early Universe that affected the balance between particles
and anti-particles, leading to the matter anti-matter asymmetry we
need to exist \cite{leptogenesis}.  This question can, in principle, be resolved if  neutrinoless double
beta decay is observed.  Because such a phenomenon will violate the
lepton number by two units, it cannot be caused if the neutrino is
different from the anti-neutrino (see \cite{murayama} and references therein).  
Many experimental proposals exist that will increase the sensitivity to such a
phenomenon dramatically over the next ten years (e.g., \cite{0nbb} and references therein).  

The crucial question cosmology can address is if a negative result from such experiments can
lead to a definitive statement about the nature of neutrinos.  
Within three generations of neutrinos, and given all neutrino
oscillation data, there are three possible mass spectra: a) degenerate, with mass 
splitting smaller than the neutrino masses, and two non-degenerate cases, b) normal hierarchy, 
with the larger mass splitting between the two more massive neutrinos and c) inverted hierarchy, 
with the smaller spitting between the two higher mass neutrinos. 
For the inverted hierarchy, a lower bound can be derived on the effective neutrino 
mass \cite{murayama}. The bound for the degenerate spectrum is stronger than for inverted
hierarchy. Unfortunately, for the normal hierarchy, one cannot obtain a similar
rigorous lower limit.  

Neutrino oscillation data have measured the neutrino squared mass
differences, which are hierarchical. Given the smallness of neutrino masses and 
the hierarchy in mass splittings, we can characterize the impact of neutrino masses  on cosmological observables and in particular on the the matter power spectrum by two parameters: the total mass $\Sigma$ and
the ratio of the largest mass splitting to the total mass, $\Delta$. One can safely neglect the impact of the solar mass splitting in cosmology. In this approach, two masses characterize the neutrino mass
spectrum, the lightest one, $m$, and the heaviest one, $M$. 

\begin{center}
\begin{figure*}[!t]
\includegraphics[width=7.5cm]{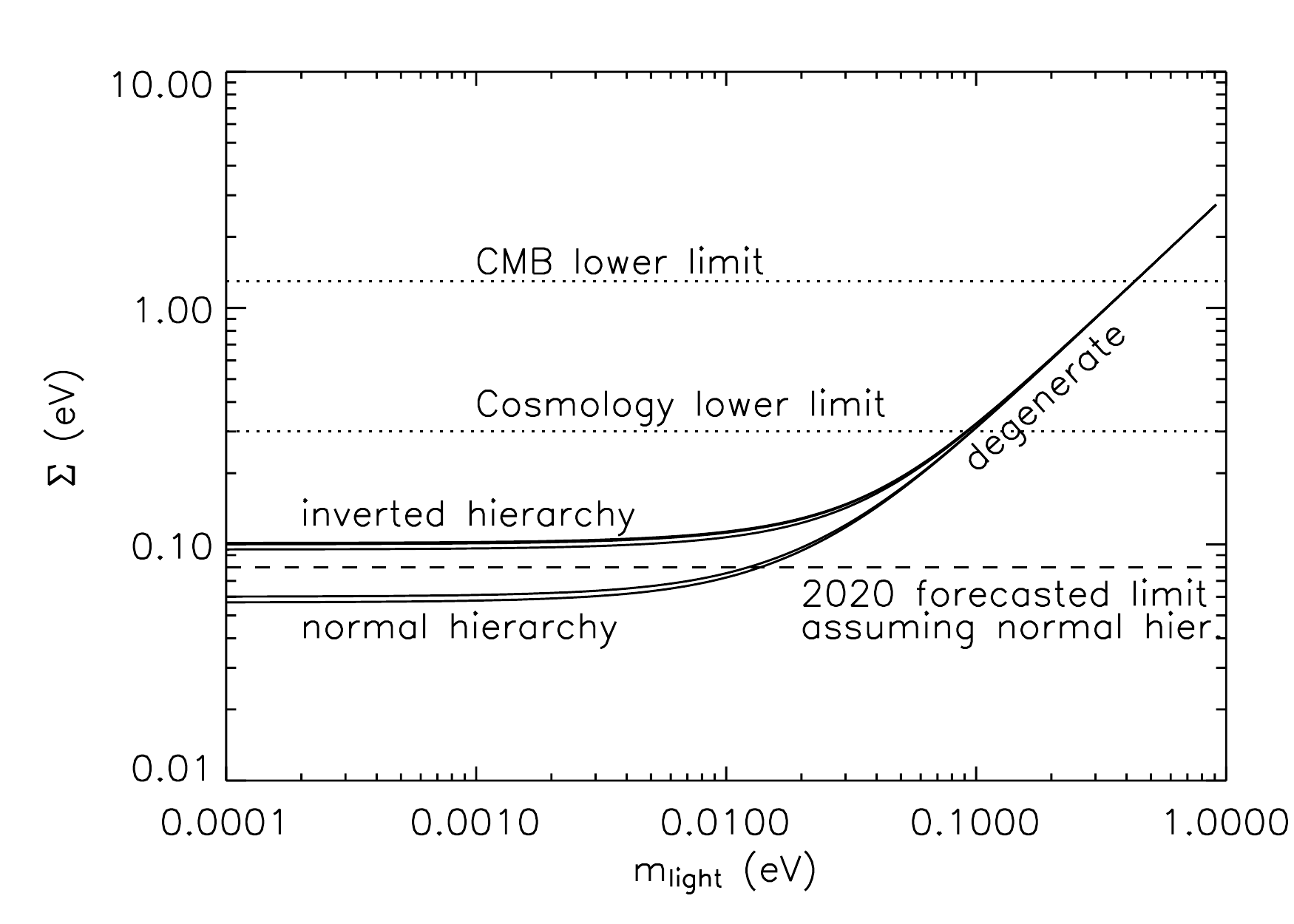}
\includegraphics[width=7.5cm]{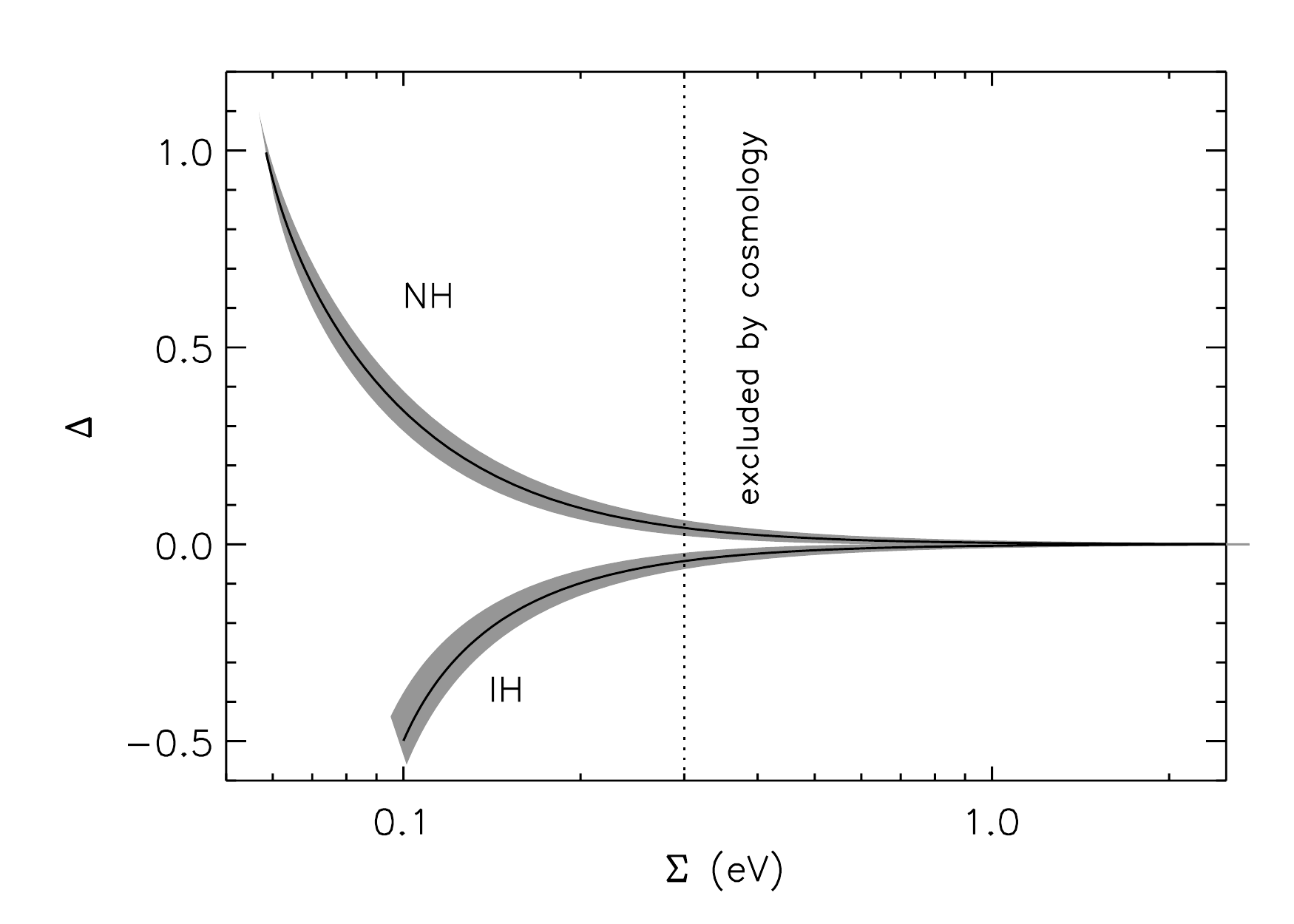}
\caption{Left: constraints from neutrino oscillations and from cosmology in the $m$-$\Sigma$ plane. Right: constraints from neutrino oscillations  (shaded regions) and from cosmology in the $\Sigma$-$\Delta$ plane. In this parameterization the sign of $\Delta$ specifies the hierarchy.} 
\label{fig:0}
\end{figure*}
\end{center}

Neutrino
oscillation data are unable to resolve whether the mass spectrum
consists in two light states with mass $m$ and a heavy one with mass $M$,
named normal hierarchy (NH) or two heavy states with mass $M$ and a
light one with mass $m$, named inverted hierarchy (IH). Near future
neutrino oscillation data may resolve the neutrino mass hierarchy if
one of the still unknown parameters that relates flavor with mass
states is not too small. On the contrary, if that mixing angle is too
small, oscillation data may be unable to solve this issue. 
Analogously, a total neutrino mass determination from cosmology 
will be able to determine the hierarchy only if the underlying 
model is normal hierarchy and $\Sigma<0.1$ eV (see e.g.,  Fig~\ref{fig:0}).
If neutrinos exist in either an inverted hierarchy or are denegerate, 
(and if the neutrinoless double beta decay signal is not seen within
the bounds determined by neutrino oscillation data), 
then the three light neutrino mass eigenstates (only) will be found 
to be Dirac particles.

Massive neutrinos affect  cosmological observations in a variety of different ways.
For example, cosmic microwave background (CMB) data alone constrain the total
neutrino 
mass $\Sigma<1.3$ eV at the 95\% confidence level \cite{Komatsu10}. 
Neutrinos with mass $< 1$eV become non-relativistic after the 
epoch of recombination probed by the CMB, thus massive neutrinos alter
matter-radiation 
equality for a fixed $\Omega_m h^2$. After neutrinos become
non-relativistic, 
their free streaming damps the small-scale  power and modifies the
shape of the matter 
power spectrum below the free-streaming length. 
Combining large-scale structure and CMB data, 
at present the sum of masses is constrained to be $\Sigma<0.3$ eV \cite{Reidnu}.
Forthcoming large-scale structure data promise to determine the
small-scale ($0.1 < k < 1$ h/Mpc) matter power spectrum
exquisitely well  and to yield errors on $\Sigma$ well below $0.1$ eV
(e.g., \cite{LSST, steen, Euclid}).

The effect of neutrino mass on the CMB is related to the
physical density of neutrinos, and therefore the mass difference
between eigenstates can be neglected. However individual neutrino
masses can have an effect on the large-scale shape of the matter power
spectrum. 
In fact, neutrinos of different masses have  different transition
redshifts from 
relativistic to non-relativistic behavior, and their individual masses
and their mass splitting change
the details of the radiation-domination to matter-domination regime. 
As a result the detailed shape of the matter power spectrum on scales
$k\sim 0.01$ $h$/Mpc is affected. 
In principle therefore a precise measurement of the matter 
power spectrum shape can give information on both the sum of the
masses 
and individual masses (and thus the hierarchy), 
even if the second effect is much smaller than the first.

We define 
the relation between the neutrino masses $m$ and $M$ and the parameters $\Sigma$ and $\Delta$ as 
\begin{eqnarray}
{\rm NH:} \, \, \,\,\,\,\,\, & \Sigma =  2m + M \,\,\,\,\,\, & \Delta=(M-m)/\Sigma \\
{\rm IH:} \, \, \,\,\,\,\,\, & \Sigma =  m + 2M \,\,\,\,\,\, & \Delta=(m-M)/\Sigma
\end{eqnarray} 
(recall that $m$ denotes  the lightest neutrino mass and $M$ the heaviest).

 In Fig~\ref{fig:0} we show the current constraints on neutrino mass
 properties in the $m$-$\Delta$ and $\Sigma$-$\Delta$ planes.
 While many different parameterizations have been proposed in the 
 literature to account for neutrino mass splitting in a cosmological
 context \cite{slosar,takada,melchiorri} 
 here we advocate using the $\Delta$ parameterization for the
 following reasons. 
 $\Delta$ changes continuously through normal, degenerate and inverted
 hierarchies; 
 $\Delta$ is positive for NH and negative for IH. Finally, see Ref.\cite{neuH},
 cosmological data are sensitive to $\Delta$ in an easily understood
 way through the largest mass splitting (i.e., the absolute value of $\Delta$), 
 while the direction of the splitting (the sign of $\Delta$) introduces 
 a sub-dominant correction to the main effect. This parameterisation
 is strictly only applicable for $\Sigma > 0$, but oscillations experiments already set $\Sigma>M > 0.05$eV.  

It is important to note that not the entire parameter space in the
$\Sigma$-$\Delta$ 
plane (or of any other parameterization of the hierarchy used in the
literature) is allowed  by particle physics
constraints and should be explored: only the regions
around the normal and inverted 
hierarchies allowed by neutrino oscillation experiments should be 
considered (see Fig~\ref{fig:0}).

To gain a physical intuition on the effect of neutrino properties 
on cosmological observables, such as the shape of the matter power
spectrum, 
it is useful to adopt the following analytical approximation, 
as described in Ref. \cite{takada}.
The matter power spectrum can be written as:
\begin{equation}
\frac{k^3 P(k;z)}{2 \pi^2} = \Delta_R^2 \frac{2 k^2}{5 H_0^2 \Omega_m^2}  D^2_{\nu} (k,z) T^2(k) \left ( \frac{k}{k_0} \right )^{(n_s-1)},
\end{equation}
where $ \Delta_R^2$ is the primordial amplitude of the fluctuations,
$n_s$ is the primordial 
power spectrum spectral slope, $T(k)$ denotes the matter transfer
function and 
$D_{\nu} (k,z)$ is the  scale-dependent linear growth function, which encloses the dependence
of $P(k)$ on non-relativistic neutrino species.

Each of the three neutrinos contributes to the neutrino mass fraction
$f_{\nu,i}$ 
where $i$ runs from $1$ to $3$,
\begin{equation}
f_{\nu,i} = \frac{\Omega_{\nu,i}}{\Omega_m} = 0.05 \left ( \frac{m_{\nu_i}}{0.658 {\rm eV}} \right ) \left ( \frac{0.14}{\Omega_m h^2} \right )
\end{equation}
and has a free-streaming scale $k_{{\rm fs},i}$,
\begin{equation}
  k_{{\rm fs},i} = 0.113  \left ( \frac{m_{\nu_i}}{1 {\rm eV}} \right )^{1/2} \left ( \frac{\Omega_m h^2}{0.14} \frac{5}{1+z}\right)^{1/2}  {\rm Mpc}^{-1}\,.
\end{equation}
Analogously, one can define the corresponding quantities 
for the combined effect of all species, by using $\Sigma$ instead of $m_{\nu_ i}$.
Since we will only distinguish between a light and a heavy eigenstate
we will have e.g., 
$f_{\nu,m}, f_{\nu,\Sigma},  k_{{\rm fs},m},  k_{{\rm fs},\Sigma}$
etc., 
where in the expression for  $f_{\nu,m}$ one should use 
the mass of the eigenstate 
(which is the mass of the individual neutrino, or twice as much 
depending on the hierarchy) while in  $k_{{\rm fs},m}$ one should use
the mass of the 
individual neutrino.

The dependence
of $P(k)$ on non-relativistic neutrino species is  in  $D_{\nu} (k,z)$, given by 
\begin{equation}
D_{\nu_i} (k,z) \propto (1-f_{\nu_i}) D(z)^{1-p_i}
\end{equation} 
where $k \gg k_{{\rm fs},i} (z)$ and $p_i = (5- \sqrt{25 - 24
  f_{\nu_i}})/4$. 
The standard linear growth function $D(z)$ fitting formula is taken
from \cite{HuEise}.  

\begin{center}
\begin{figure*}
\includegraphics[trim= 0mm 120mm 0mm 0mm, clip=true, width=7.8cm]{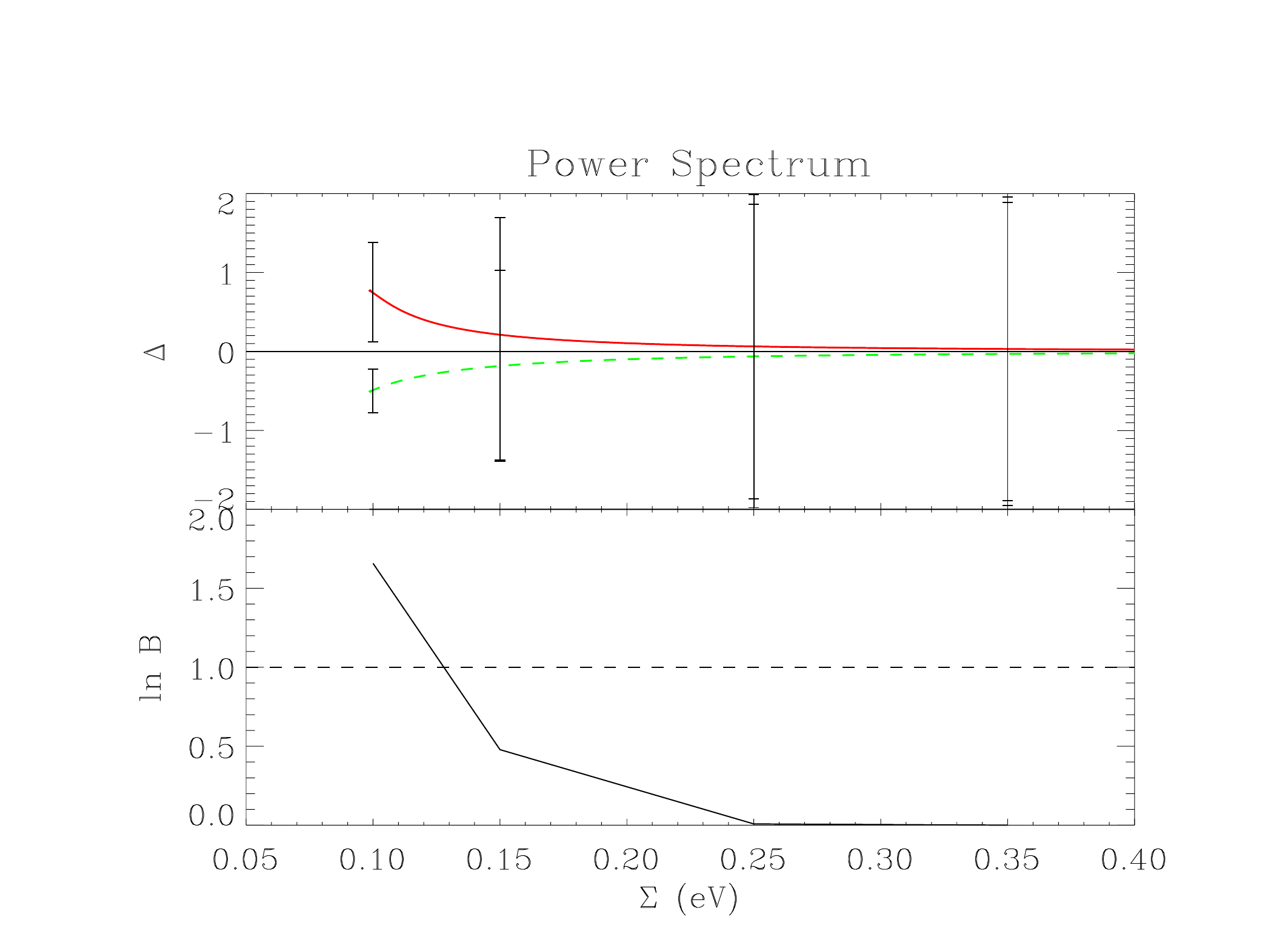}
\includegraphics[trim= 0mm 120mm 0mm 0mm, clip=true, width=7.8cm]{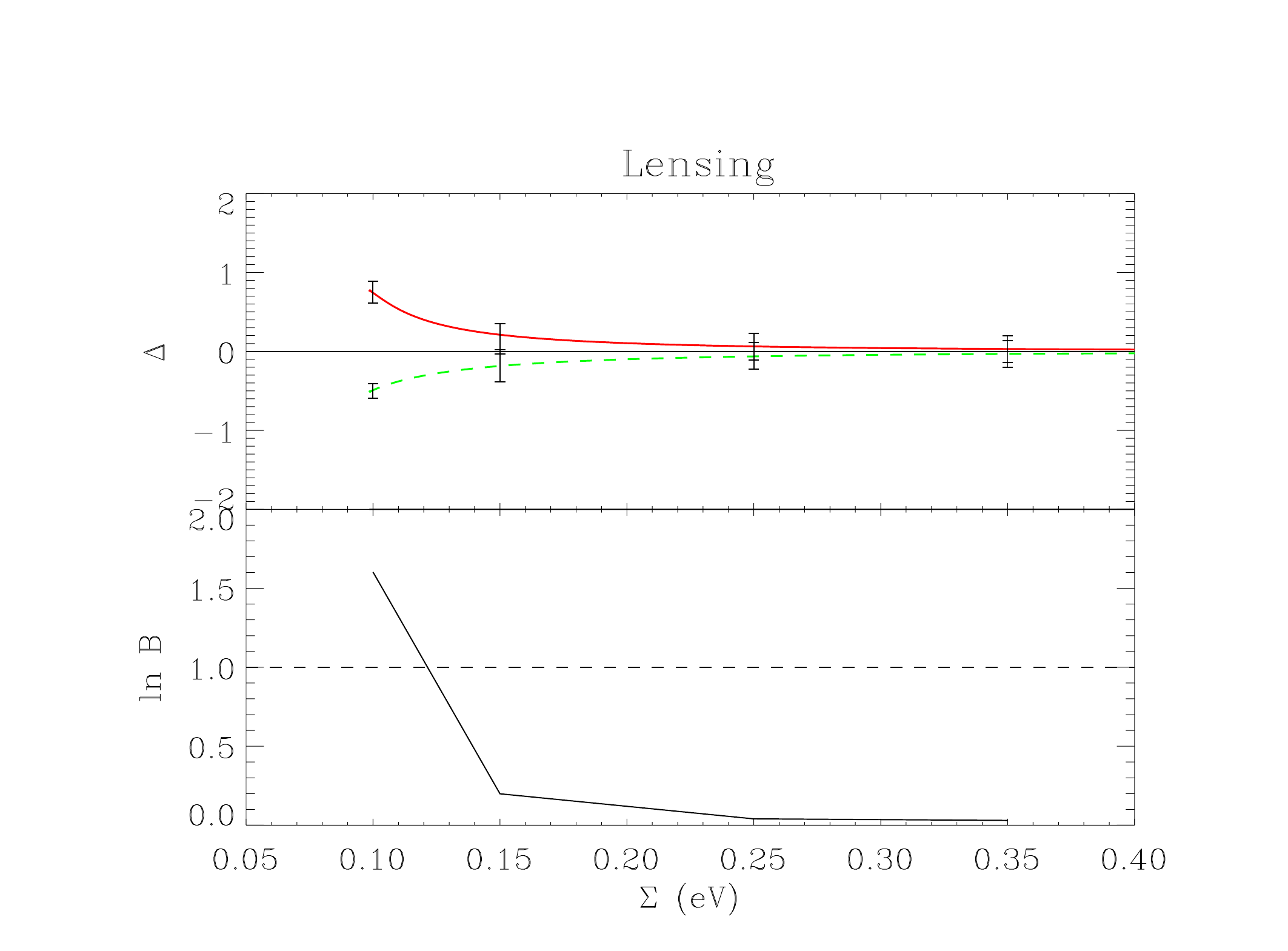}
\caption{LSS (left) and Weak Lensing (right) 
    forecasts for neutrino mass parameters $\Sigma$ and
    $\Delta$. We assume the LSS survey
    consists of a comoving volume of $600$ Gpc$^3$ at $z=2$ and $2000$ Gpc$^3$ at
    $z=5$. The Weak Lensing survey covers $40$,$000$ sq. deg. with a
    median redshift of $3.0$ and a number density of $150$ galaxies per
    square arcminute.
    Several
    fiducial models ($\Sigma$,$\Delta$) were used to derive by Fisher matrix
    approach the expected errors. The upper panel shows the $1$-$\sigma$
    errors on $\Delta$ and $\Sigma$, the errors in $\Sigma$ are so small
    that are barely visible. 
    The lower panel shows the expected evidence ratio between the normal and 
    inverted constraints as a function of neutrino mass. The dashed line
    shows the $\ln B=1$ level: in Jeffrey's scale $\ln B<1$  is
    `inconclusive' evidence, 
    and $1<\ln B<2.5$ corresponds to `substantial' evidence.} 
\label{fig:gal}
\end{figure*}
\end{center}

\begin{center}
\begin{figure*}
\includegraphics[width=\textwidth]{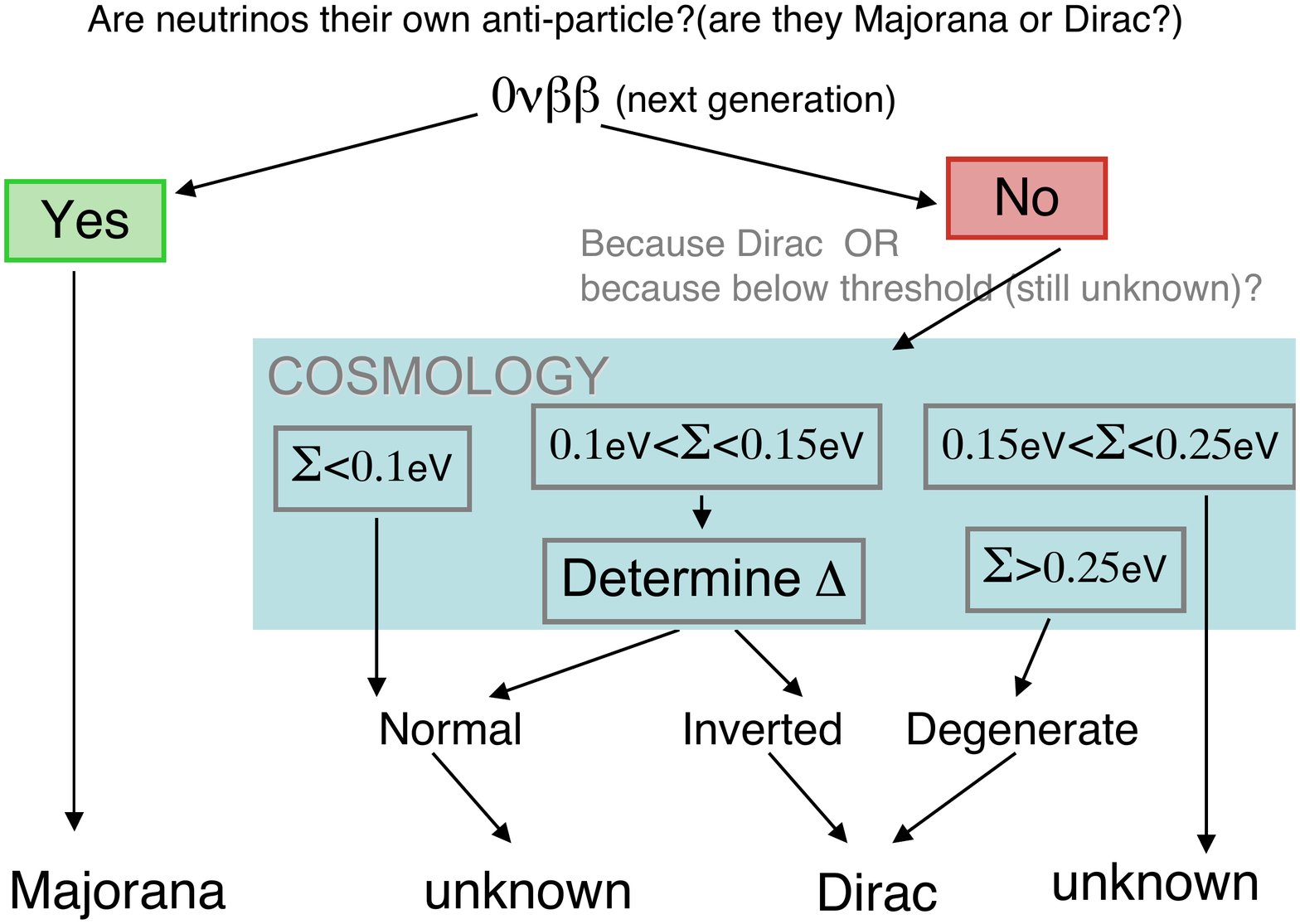}
\caption{Role of cosmology in determining the nature of neutrino mass.  Future  neutrinoless double beta decay ($0\nu \beta \beta$)  experiments and future cosmological surveys will be highly complementary in addressing the question of whether neutrinos are Dirac or Majorana particles. Next generation  means  near future experiments whose goal is to reach a sensitivity to the neutrinoless double beta decay effective mass of $0.01$ eV. We can still find two small windows where this combination of experiments will not be able to give a definite answer,  but this region is much reduced by combining  $0\nu \beta \beta$ and cosmological observations.} 
\label{fig:flowchart}
\end{figure*}
\end{center}

In summary there are three qualitatively different regimes in $k$-space
that are introduced by the neutrino mass splitting
\begin{eqnarray}
 D_{\nu} (k,z)=& D(k,z)  & \,\,\,\,\,\,\, k < k_{{\rm fs},m} \\
 D_{\nu} (k,z)=& (1\!-\!f_{\nu,m})D(z)^{(1-p_m)} & \,\,\,\,\,k_{{\rm fs},m}\!<\!k\!<k_{{\rm fs},\Sigma}\\
 D_{\nu} (k,z)=& (1-f_{\nu,\Sigma})D(z)^{(1-p_{\Sigma})} &\,\,\,\,\,\,\, k>k_{{\rm fs},\Sigma}\,,
 \end{eqnarray}
where the subscript $m$ refers to the light neutrino eigenstate and $\Sigma$ to the sum of all masses.
 
This description is, however, incomplete: the transitions between the
three 
regimes is done sharply in $k$ while in reality the change is very
smooth. 
In addition, the individual masses change the details of the
matter-radiation 
transition which (keeping all other parameters fixed) adds an
additional effect at scales $k>k_{{\rm fs},\Sigma}$.

In order to explore what constraints can be placed on $\Delta$ and
$\Sigma$ for a given survey set-up we can use a Fisher matrix approach. 
The elements of ${\bf F}$, the  Fisher information matrix \cite{Fisher}, are given by 
\begin{equation}
  F_{\theta \gamma}=-2 \left \langle \frac{\partial^2 \ln L}{\partial \theta \partial \gamma}\right \rangle
\end{equation}
where  $\theta$  and $\gamma$ denote cosmological parameters 
(and the Fisher matrix element's indices) and $L$ denotes the likelihood of the data given the model.
Marginalised errors on a parameter are computed
as $\sigma^2 (\theta) = ({\bf F}^{-1})_{\theta \theta}$ .
We can also
calculate expected Bayesian evidence for cosmological parameters using
the approach of Ref.~\cite{HKV,TK10}. In the case that
we are considering we use the formula from \cite{TK10} for the
expectation value of the evidence, in this case the expected Bayes
factor is simply the log of ratio of the Fisher determinants. 
 
Following Ref.~\cite{Seo} the Fisher matrix for the galaxy power spectrum is
\begin{equation}
F_{\theta \gamma} = \frac{V_s}{8 \pi^2} \int_{-1}^{1} d\mu
\int_{k_{\rm min}}^{k_{\rm max}}  k^2 dk N \frac{\partial \ln P(k,\mu)}{\partial \theta} \frac{\partial \ln P(k,\mu)}{\partial \gamma}  
\label{eq:fisher}
\end{equation} 
with $N=[nP(k,\mu)/(nP(k,\mu)+1)]^2$ and $V_s$ is the volume of the
survey. The integration over the projected angle along the light of
sight
\footnote{As it is customary, $\mu$ denotes the cosine of the angle
  with respect to the line of sight.}
$\mu$ is analytical and
the maximum and minimum  wavenumbers allowed depend
on the survey characteristics with the constraint that $k_{\rm max}$ must be in the
linear regime. 
The derivatives are computed at the fiducial model chosen.

The degeneracies between $\Sigma$ and $\Delta$ are small, and the very
small constraint on $\Sigma$ results in the constraints being
effectively un-correlated in the $\Sigma$-$\Delta$ plane. 
We note that the constraints on $\Delta$ around the IH and
NH peaks are tighter for weak lensing than LSS, this is due to
lensing providing constraints on both the geometry and the growth of
structure, which provides a smaller raw constraint and a more
orthogonal constraint to the CMB resulting in smaller
errors. Interestingly, even though the weak lensing constraints on
$\Delta$ are smaller than for the power spectrum, the evidence ratio
is comparable (see Fig.~\ref{fig:gal}), because, 
due to the multi-dimensional degeneracy directions, 
a naive correspondence between error-bars and evidence is not
applicable (it is to a first approximation the difference between the two error
bars that is important).

Note that the evidence
calculation explicitly assumes two isolated peaks, and so is only
applicable when the fiducial points are seperated by multiple-sigma. 
As a result of this, the evidence calculations may be slightly optimisic
for large masses. However, for $\Sigma <0.2$ eV, the $\chi^2$ difference between the two minima becomes noticeable   as well as the shift between the location of one of the two minima  and the central $\Delta$ value for the oscillations experiments (which induces an additional $\chi^2$ difference). While this information  is not fully accounted for in a Bayesian approach to forecasting  the evidence, it may be included at the moment of analyzing the data, using different approaches such as the likelihood ratio, and may slightly improve the significance for  the hierarchy determination.

While we have used the oscillation results to center the Fisher and evidence calculations on the NH and IH, combining the oscillation experiments constraints will not improve the evidence; in fact,  oscillation experiments give symmetric errors  on $\Delta$ (i.e. they do not depend on the sign of $\Delta$). The final scheme is shown in Fig.~\ref{fig:flowchart} which illustrates how the hierarchy can be determine in future cosmology experiments.

Better perspectives at measuring the neutrino hierarchy from the sky can be attained by exploiting the non-linear part of the power spectrum of galaxies as was shown by Ref.~\cite{wagnernu}. Non-linearities enhance the dependence of the power spectrum on the different neutrino hierarchies, thus making the observational signature more pronounced. 
If all other cosmological parameters are known (including the sum of neutrino masses $\Sigma$), the two hierarchies can  be distinguished with confidence, as illustrated in Fig.~\ref{fig:deltachisq} as function of the maximum $k$ considered, making the effect potentially measurable. We have assumed an effective volume of $1\,({\rm Gpc}/h)^3$ at $z=0$ (red lines) and  $10\,({\rm Gpc}/h)^3$ at $z=1$ (blue lines).\footnote{These volumes roughly correspond  to the volume out to $z=0.5$ and  between $z=0.5$ and $z=1.5$ in $1/10$ of the sky respectively in a standard $\Lambda$CDM universe.} Whether degeneracies with other cosmological parameters and systematic effects (galaxy bias, baryonic physics, observational limitations etc.) will cancel the detectability of the effect remains to be explored.
 
Cosmology has the potential of determining the neutrino hierarchy in the interesting window $\Sigma \gtrsim 0.1$eV.  Signal-to-noise estimates done using  linear theory predictions  indicated that if $\Sigma$ happens to be in  the window $0.15{\rm eV}<\Sigma < 0.25$eV, cosmology could not help determine the hierarchy and thus the nature of neutrino masses \citep{neuH}, leaving an important gap in our knowledge of neutrino properties.  The fact that non-linearities enhance the effect compared to the linear prediction, will potentially enable cosmology to determine the hierarchy in a wider $\Sigma$ range and  possibly  close the gap. 

As an aside and already noted in Ref.~\cite{neuH}, cosmology is  more sensitive to $|\Delta |$ than to its sign: a measurement of $|\Delta |$ in agreement with that predicted by oscillations experiments for the measured $\Sigma$ would provide a convincing consistency check for the total neutrino mass constraint from cosmology.

\begin{figure}[tb]
\begin{center}
\includegraphics[trim= 20mm 20mm 5mm 10mm, clip=true ,angle=0,width=0.95\textwidth]{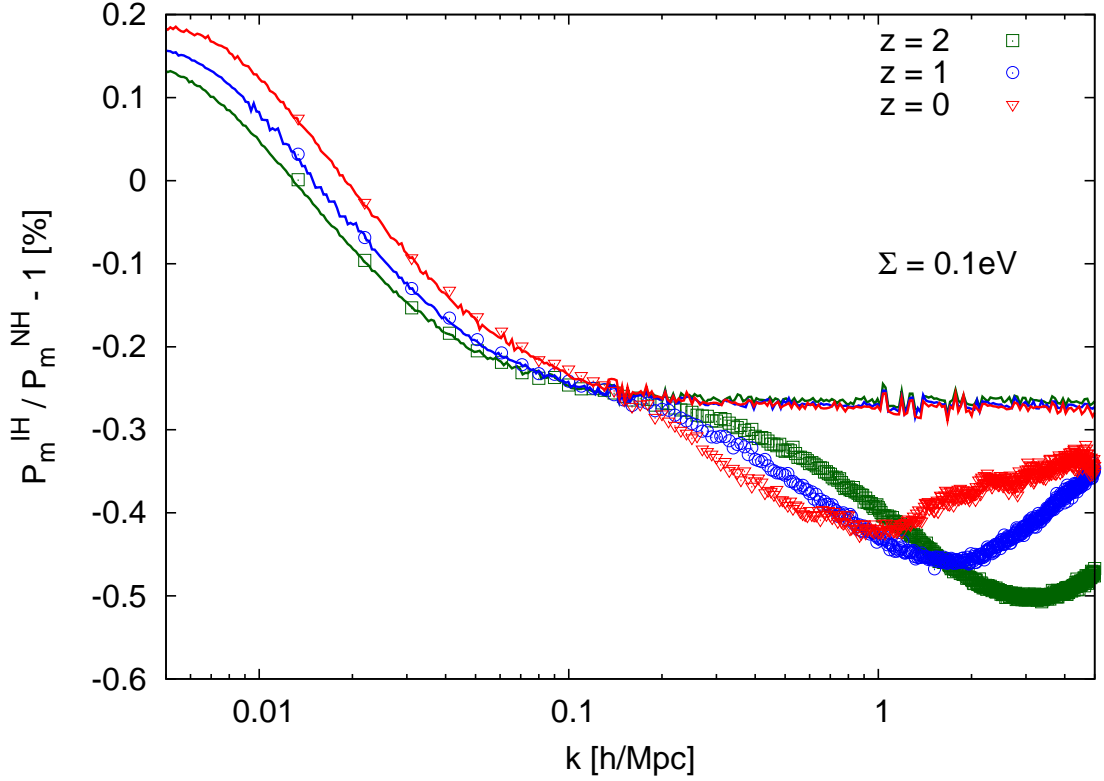}
\end{center}
\caption
{Fractional difference in the total matter power spectrum (CDM+baryons+massive neutrinos) of the inverted hierarchy and normal hierarchy run (the  sum of neutrino masses is kept fixed and only the mass splitting is varied). Note that also in this case, non-linearities enhance the effect on mildly non-linear scales compared to linear theory predictions (solid lines).
}
\label{fig:Pk_ratio_NH_IH}
\end{figure}

\begin{figure}[htb]
\begin{center}
\includegraphics[trim= 20mm 25mm 0mm 0mm, clip=true ,angle=0,width=0.95\textwidth]{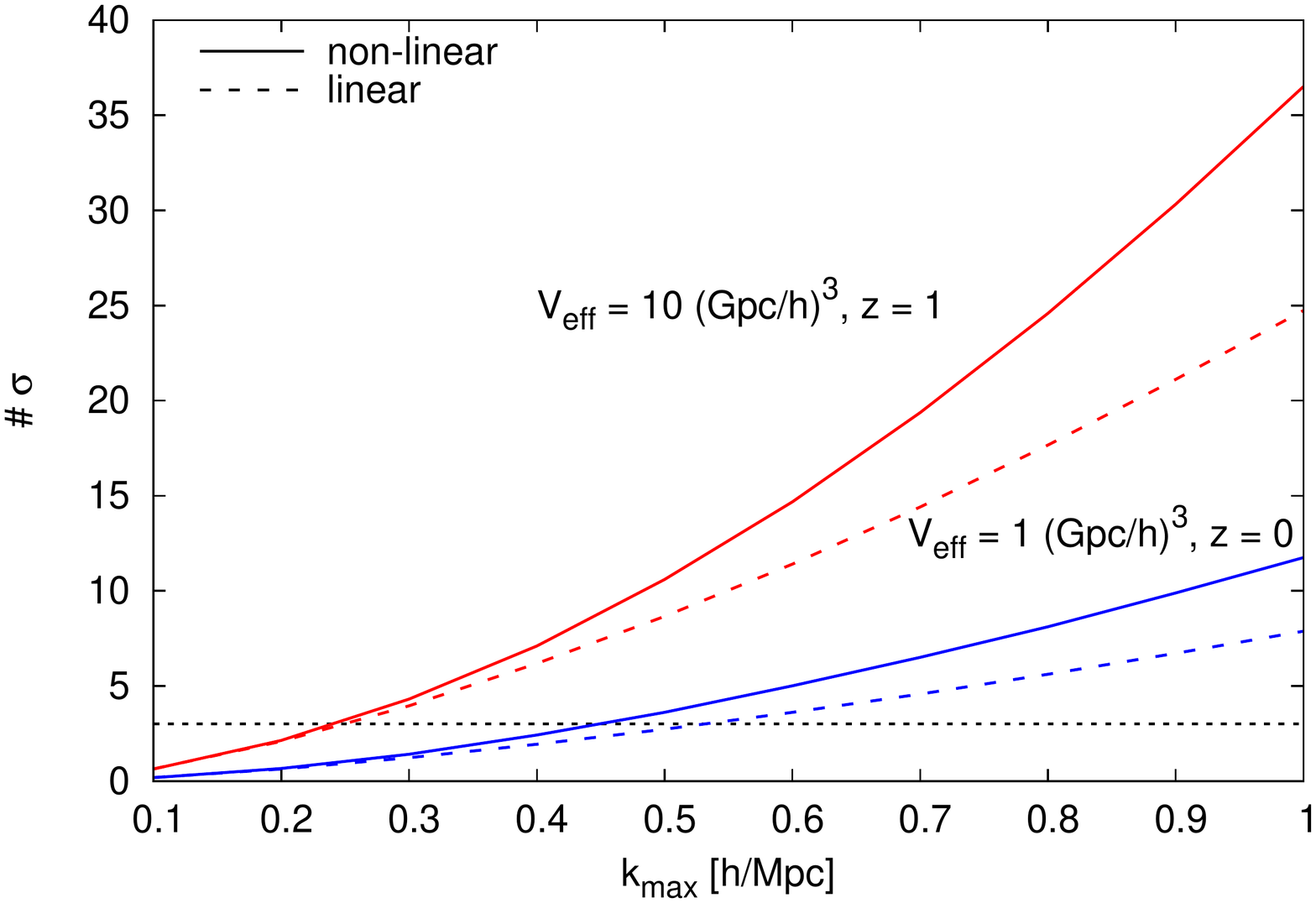}
\end{center}
\caption{Forecast of the number of sigmas separating the two hierarchies for $\Sigma=0.1$eV as a function of the maximum $k$ vector considered and for two effective volumes --$1$ and $10\, ({\rm Gpc}/h)^3$-- at $z=0$ and $z=1$ respectively. It is  assumed that all other cosmological parameters are known. The horizontal dotted line indicates the $3-\sigma$  level.}
\label{fig:deltachisq}
\end{figure}

\section{Using astronomical distances to constraint beyond the standard model physics}

Cosmological observations provide constraints on different distance measures: luminosity distance (as provided e.g., by supernovae), angular diameter distance (as provided e.g., by baryon acoustic oscillations) and even on the expansion rate or the Hubble parameter as a function of redshift $z$. 
Both luminosity distance and angular diameter distance are functions
of the Hubble parameter. While combining these measurements helps to
break parameter degeneracies and constrain cosmological parameters,
comparing them helps to constrain possible deviations from the assumptions
underlying the standard cosmological model (e.g. isotropy), or to directly constrain 
physics beyond the standard model of particle physics (e.g. couplings of photons 
to scalar or pseudo-scalar matter).

The Etherington relation \cite{Etherington1} implies that,  in a cosmology based on a metric theory of gravity,  distance measures are unique: the luminosity distance is $(1+z)^2$ times the  angular diameter distance. This is valid in any cosmological background where photons travel on null geodesics and  where, crucially, photon number is conserved.

There are several scenarios in which the Etherington relation would be violated: for instance we can have  deviations from a metric theory of gravity, photons not traveling along unique null geodesics,  variations of fundamental constants, etc.
Here we want to restrict our attention to violations of the Etherington relation arising from the violation of photon conservation. 

A change in the photon flux during propagation towards the Earth  will affect the Supernovae (SNe) luminosity distance measures but not the determinations of the angular diameter distance.  
Photon conservation can be violated by simple astrophysical effects or by exotic physics.
Amongst the former we find, for instance, attenuation due to interstellar dust, gas and/or plasmas. Most known sources of attenuation are expected to be clustered and can be typically constrained down to the 0.1\% level \cite{Menard, Bovy}. Unclustered sources of attenuation are however much more difficult to constrain.  For example, gray dust \cite{Aguirre}  has been invoked to explain the observed dimming of Type Ia Supernovae without  resorting to cosmic acceleration.

More exotic sources of photon conservation violation involve a
coupling of photons to particles  beyond the standard model of
particle physics.  Such couplings would mean that, while passing
through the intergalactic medium, a photon could disappear --or even (re)appear!-- interacting with such  exotic particles, modifying the apparent luminosity of sources.  
Here we consider the mixing of photons with scalars, known as axion-like
particles, and the possibility of mini-charged particles which have a
tiny, and unquantised electric charge.  
A recent review \cite{Jaeckel:2010ni} highlights the rich phenomenology of these weakly-interacting-sub-eV-particles (WISPs), whose effects have been searched for in a
number of laboratory experiments and astronomical observations. 
In particular, the implications of this particles on the SN luminosity have been described in a number of publications~\cite{Csaki2,Mortsellaxions,Mirizzi:2006zy,Burrage:2007ew,Ahlers:2009kh}. 

One of the most interesting features of these models is that the exotic opacity involved could  in principle ``mimic" the value of a non-zero cosmological constant inferred from SNe measurements. However, this possibility can already be excluded (at least in the simplest WISP models) by the absence of distortions in the CMB or the spectra of quasars for axion-like-particles, and by arguments of stellar evolution in the case of mini-charged particles. 

\subsection{Calculating cosmic opacity}

In reference \cite{AVJ}, the authors use Type Ia SN brightness 
data (namely the SCP Union 2008 compilation \cite{Union}) in combination
with measurements of cosmic expansion $H(z)$ from differential aging 
of luminous red galaxies (LRGs) \cite{JVST,svj} to obtain constraints 
on non-trivial opacity, at cosmological scales.  The basic idea is to study 
possible violations from the ``Etherington relation'' \cite{Etherington1}, the 
distance duality between luminosity distance, $d_L$, and angular diameter 
distance, $d_A$:
\begin{equation}\label{Etherington}
d_L(z)=(1+z)^2 d_A(z)\, .
\end{equation}
This identity depends only on photon number conservation and local Lorentz 
invariance.  It holds for general metric theories of gravity, where photons travel 
along unique null geodesics.  Since Lorentz violation is strongly constrained for 
the low energies corresponding to optical observations \cite{Kostelecky:2008ts}, 
the study of possible violations of Eq.~(\ref{Etherington}) through SN 
observations directly constrains photon number violation.  Any such 
systematic violations can then be interpreted as an opacity effect in the 
observed luminosity distance, parametrised through a generic opacity 
parameter, $\tau(z)$, as:
\begin{equation}\label{tau} 
d^2_{L,obs}=d^2_{L,true} e^{\tau(z)} \, .
\end{equation}  

Note that this  ``opacity'' can have in principle both signs. In 
other words, this parametrisation also allows for apparent \emph{brightening} 
of light sources, as would be the case, for example, if exotic particles were also emitted 
from the source and converted into photons along the line of sight \cite{Burrage:2007ew}. 
From Eq.~(\ref{tau}) it is clear that the inferred distance 
moduli for the observed SNe picks an extra term which is linear in $\tau(z)$:
\begin{equation}\label{DMs}
 DM_{obs}(z)=DM_{true}(z)+2.5[\log e] \tau(z) \, .
\end{equation}  

On the other hand, one can also use other determinations of distance 
measures, which are independent of $\tau$, to constrain possible deviations
from Eq.~(\ref{Etherington}).  This approach was initiated in reference 
\cite{More} (see also \cite{bassettkunz1,bassettkunz2,Uzan,LazNesPer} for related earlier 
work) where the authors used measurements \cite{PercivalBAO} of the baryon acoustic oscillation (BAO) scale at two redshifts, namely $z=0.20$ and $z=0.35$, to obtain a parameterization-independent upper-bound for the difference in  opacity between these two redshifts, 
$\Delta\tau<0.13$ at 95\% confidence.  In reference \cite{AVJ} this constraint was improved 
(and also extended over 
a wider redshift range, but for a  general parameterised form for $\tau$) by using, instead of measurements of the BAO scale at these two redshifts, measurements of cosmic expansion $H(z)$ 
from differential aging of LRGs at redshifts $z < 2$.  This method of distance determination 
relies on the detailed shapes of galaxy spectra but not on galaxy luminosities, so 
it is independent of $\tau$.  

In particular, the authors introduced a parameter $\epsilon$ to study deviations 
from the Etherington relation of the form:
\begin{equation}\label{epsilon}
d_L(z) = d_A(z) (1+z)^{2+\epsilon} \, ,
\end{equation} 
and constrained this parameter to be 
$\epsilon=-0.01^{+0.08}_{-0.09}$ at 95\% confidence.
Restricted to the redshift range $0.2<z<0.35$, where $\tau(z)=2\epsilon z + 
{\cal O}(\epsilon z^2)$, this corresponds to $\Delta\tau<0.02$ at 95\% confidence. 
Below, we will apply similar constraints on different parametrisations
of $\tau$ which correspond to particular models of exotic matter-photon coupling, namely 
 axion-like particles (ALPs), chameleons, and mini-charged particles (MCPs).  

Before moving to these models, we briefly update the above constraint on $\epsilon$ 
using the latest $H(z)$ data \cite{sternjim}, which include two extra data points at 
redshifts $z=0.48$ and $z=0.9$, as well as the latest determination of $H_0$ 
\cite{Riess09}.  Even though the addition of these two extra data points alone 
significantly improves the constraints of reference \cite{AVJ}, the effect of $H_0$ 
is also quite significant, because it acts as an overall scale in the distance measures, 
which is marginalised over a Gaussian prior, and the measurement error in this 
determination is about half of that of the HST Key Project determination 
\cite{HSTKey} used in \cite{AVJ}.  
 
\begin{figure}[h]
  \begin{center}
    \includegraphics[height=2.6in,width=2.8in]{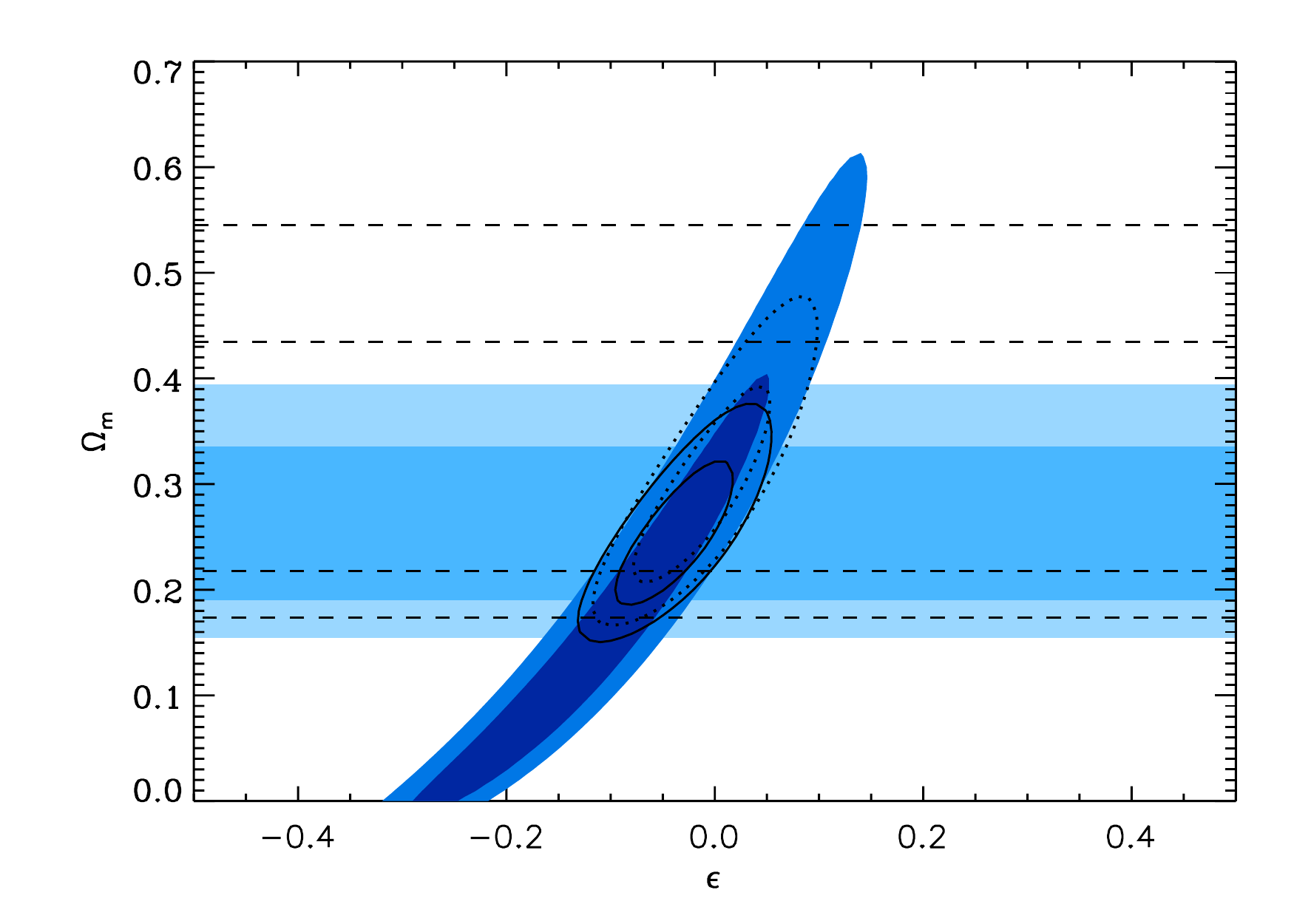}
    \includegraphics[height=2.6in,width=2.8in]{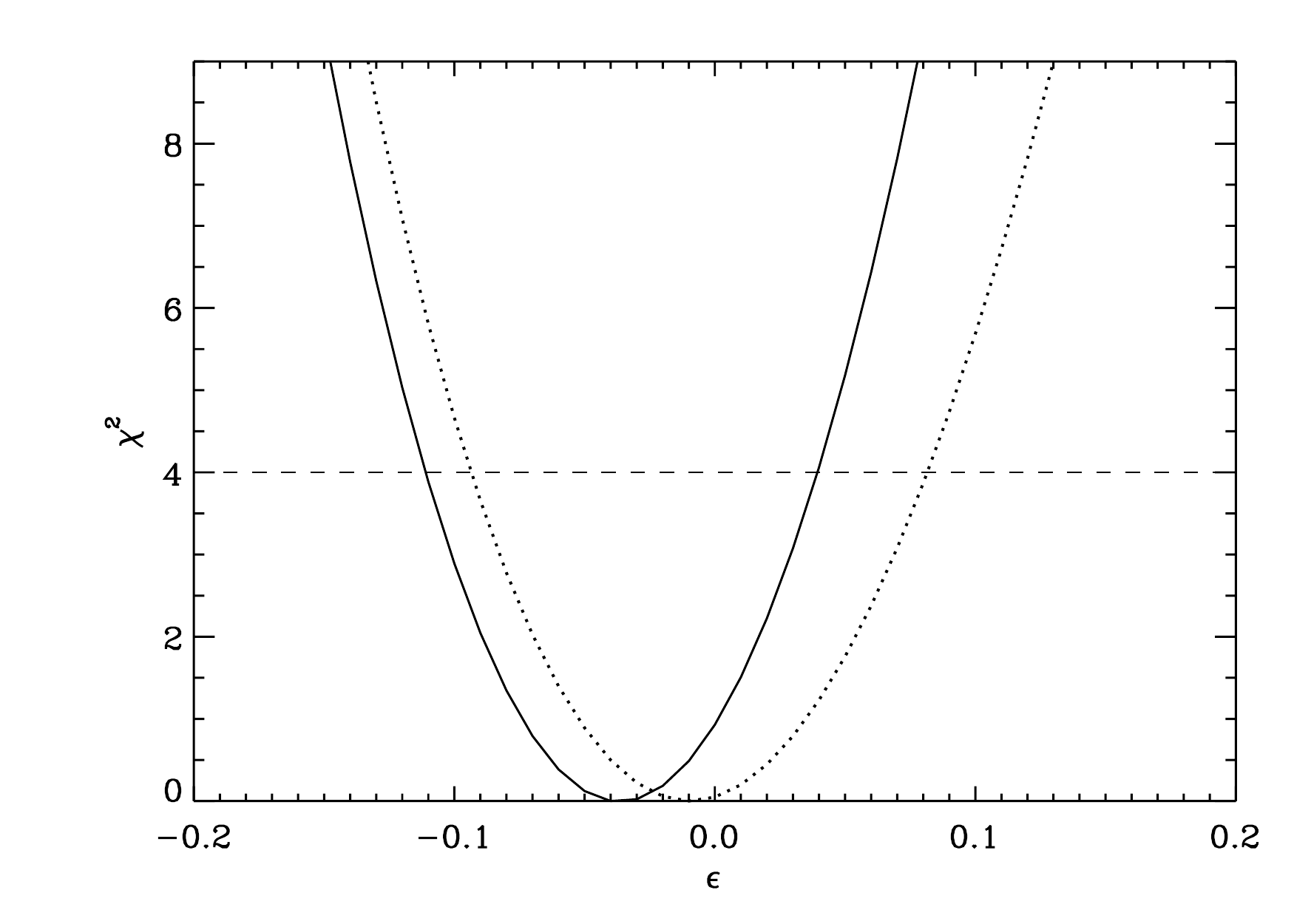} 
    \caption{\label{epsilon_constrs}  Updated constraints of reference 
                  \cite{AVJ}, using $H(z)$ data \cite{sternjim} and the 
                  Riess et al. determination of $H_0$ \cite{Riess09} in combination 
                  with the SCP Union 2008 SN Ia compilation.  
                  {\it Left:} Two-parameter constraints on the $\epsilon-\Omega_m$ 
                  plane.  Darker blue contours correspond to 68\% and 95\% 
                  confidence levels obtained from SN data alone, lighter blue 
                  contours are for $H(z)$ data, and solid line transparent contours 
                  are for joint SN+$H(z)$.  Previous $H(z)$ and joint SN+$H(z)$
                  from \cite{AVJ} are shown in dashed and dotted lines respectively.  
                  {\it Right:} One-parameter joint constraints on $\epsilon$ for the 
                  current analysis (solid line) and that of reference \cite{AVJ} (dotted
                  line). The dashed line shows the 95\% confidence level, $\Delta\chi^2=2$.}
  \end{center}
\end{figure}

Fig. \ref{epsilon_constrs} shows the updated constraints obtained using the 
above data in combination with the SCP Union 2008 Compilation \cite{Union} of type Ia 
Supernova data, compared to the previous constraints of reference \cite{AVJ}.  On the left, the darker blue contours correspond to the (two-parameter) 68\% 
and 95\% joint confidence levels obtained from SN data alone, while lighter blue 
contours are the corresponding confidence levels for $H(z)$ data.  Solid-line 
transparent contours are for joint SN+$H(z)$ data.  For comparison 
we also show the previous $H(z)$ and  SN+$H(z)$ contours in dotted
and dashed lines respectively.  On the right we show one-parameter  (marginalized over 
all other parameters) constraints on $\epsilon$, again for the current analysis (solid line) 
and for that of reference \cite{AVJ} (dotted).  For the reader familiar with Bayesian 
methods, this plot corresponds to the posterior
\begin{equation}\label{post_prob}
P(\epsilon|{\rm S},{\rm E})=\int_{\Omega_m}\int_{H_0} P(\Omega_m,H_0|{\rm E})
P(\epsilon,\Omega_m,H_0|{\rm S}) \, {\rm d}\Omega_m {\rm d}H_0 \,,
\end{equation}
where $P(\Omega_m,H_0|{\rm E})$ and $P(\epsilon,\Omega_m,H_0|{\rm S})$ are
the posterior probabilities for the corresponding model parameters, given
the $H(z)$ ({\bf E}xpansion) and SN ($\bf S$upernovae) data respectively.
These are given by the likelihoods of the two data sets in the model parameters,
assuming Gaussian errors and using flat priors on all three parameters.  In 
particular, we have taken $\epsilon\in [-0.5,0.5]$, $\Omega_m\in [0,1]$ and 
$H_0\in [74.2-3\times 3.6, 74.2+3\times 3.6]$ (Riess et. al. \cite{Riess09}), 
all spaced uniformly over the relevant intervals, in a flat $\Lambda$CDM model.  
Similarly, the solid line transparent contours on the left plot of 
Fig. \ref{epsilon_constrs} correspond to taking only the integral over $H_0$ 
in the right hand side of Eq.~(\ref{post_prob}), yielding, therefore, the 
two-parameter posterior $P(\epsilon,\Omega_m |{\rm S},{\rm E})$.    

As seen in Fig. \ref{epsilon_constrs}, the improvement in these constraints is
significant.  The new result on $\epsilon$, marginalised over all other parameters, 
is $\epsilon=-0.04^{+0.08}_{-0.07}$ at 95\% confidence, which for redshifts between 
$0.2$ and $0.35$ (currently probed by BAO data), corresponds to a transparency (i.e., $\tau\ge 0$) 
bound $\Delta\tau<0.012$, a factor of two tighter than the result in reference \cite{AVJ}\footnote{Note that the data slightly favour negative $\epsilon$ (thus the much stronger constraint on a positive $\Delta\tau$), but 
only at $ < $ 1-$\sigma$ level.}.  We now move on to study 
more general parametrisations of cosmic opacity, tailored for particular models of exotic 
matter coupled to photons.

\subsection{\label{cham}Axion-like Particles and Chameleons}
New scalar or pseudo scalar particles from physics beyond the standard model, here denoted as 
$\phi$, may couple to photons through
\be\label{Lscalar}
\mathcal{L}_{scalar}=\frac{1}{4M}F_{\mu\nu}F^{\mu\nu}\phi
\ee 
and
\be\label{Lpseudo}
\mathcal{L}_{pseudo-scalar}=\frac{1}{8M}\epsilon_{\mu\nu\lambda\rho}F^{\mu\nu}F^{\lambda\rho}\phi
\ee
where $M$ is the energy scale of the coupling (another widely used notation is 
$g_{\phi\gamma}=1/M$), $F_{\mu\nu}$ the 
electromagnetic field strength and $\epsilon_{\mu\nu\lambda\rho}$ the Levi-Civita
symbol in four dimensions.  
Such fields are collectively known as Axion-Like Particles (ALPs), as a coupling of the form (\ref{Lpseudo}) arises for the 
axion introduced by Peccei and Quinn (PQ) to solve the strong CP problem \cite{Peccei:1977hh}. 
Interestingly, these fields also arise naturally in string theory (for a review see \cite{Svrcek:2006yi}). 

Axions, or axion-like-particles, can arise from field theoretic extensions of the standard model as Goldstone bosons when a global shift symmetry, present in the high energy sector, is spontaneously broken. 
In the PQ axion case, this symmetry is colour anomalous and the explicit breaking makes the axion pick up a small mass.  This mass is, up to a model-independent constant, proportional to the coupling (\ref{Lpseudo}).
For a generic ALP, however, the mass is in principle independent of the strength of its coupling, 
and in particular can be zero if the releted shift symmetry remains intact.  That is, for instance, 
the case of Arions~\cite{Anselm:1981aw}, the orthogonal combination of the PQ axion, 
if there are \emph{two} independent colour anomalous shift symmetries.  

Chameleon scalar fields are another very interesting type of ALPs~\cite{Brax:2009ey}. 
They were originally invoked in \cite{Khoury:2003aq,Khoury:2003rn} to explain the current accelerated expansion of the Universe with a quintessence field which can couple to matter without
giving rise to large fifth forces or unacceptable violations of the
weak equivalence principle.  
The chameleon achieves this because its mass depends on the local energy density. 
The environmental dependence of the mass of the chameleon means that it
avoids many of the constraints on the strength of the coupling, which normally
apply to standard scalar and pseudo-scalar fields as they are derived
from physics in dense environments. For a more detailed discussion see \cite{Burrage:2008ii}.  
The cosmology of the chameleon was explored in detail in~\cite{Brax:2004qh}, the possibility of the chameleon coupling to photons was first discussed in~\cite{Brax:2007ak} and such a coupling was 
shown to be generic in~\cite{Brax:2009ey}.

The Lagrangian terms given above mean that ALPs can affect the
propagation of photons; in particular, if photons traverse a
magnetic field there is a non-zero probability that they will
oscillate into ALPs \cite{Raffelt:1987im}.   Notice however that only
photons polarized perpendicular (parallel) to the magnetic field mix
with scalar (pseudo-scalar) particles. Therefore, the interactions between photons 
and ALPs in the presence of a magnetic field not only imply that photon number 
is not conserved, but can also alter the polarization of the light beam.
Both effects have been exploited in many searches for ALPs both in the laboratory and in
astronomical observations, see~\cite{Jaeckel:2010ni} for a recent review.

\begin{figure}[h]
  \begin{center}
    \includegraphics[height=1.9in,width=2in]{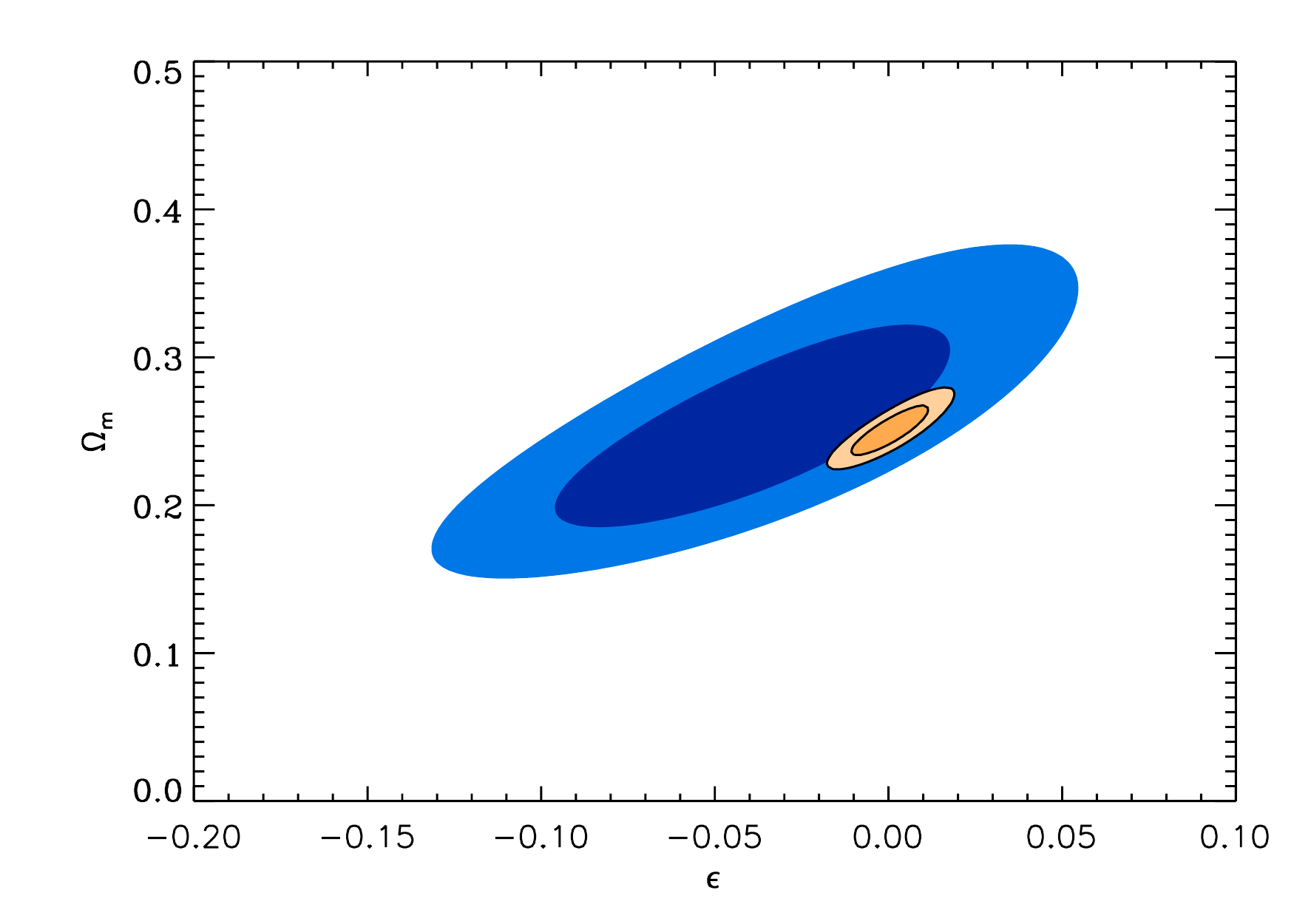} 
    \includegraphics[height=1.9in,width=2in]{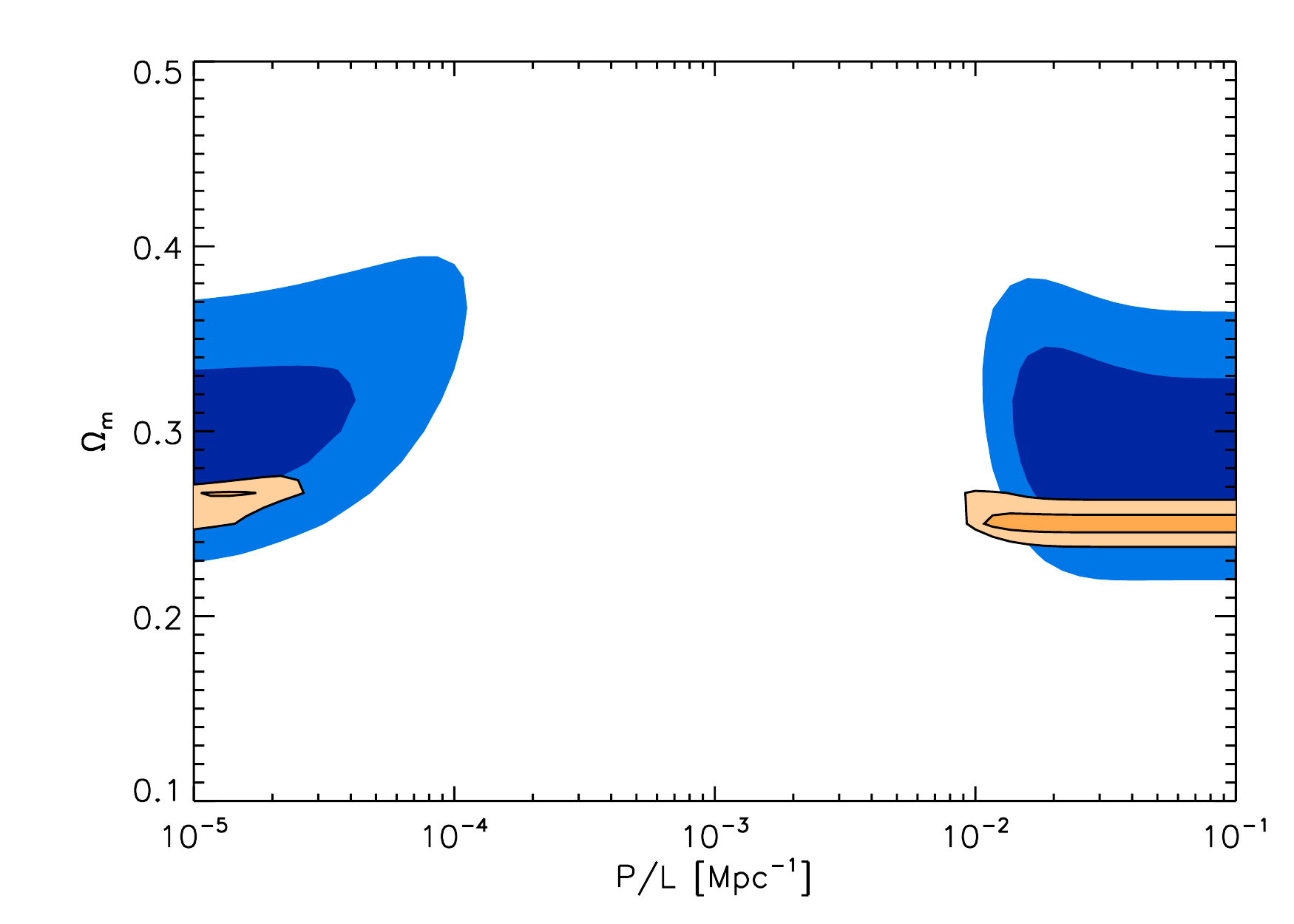}\\
    \includegraphics[height=1.9in,width=2in]{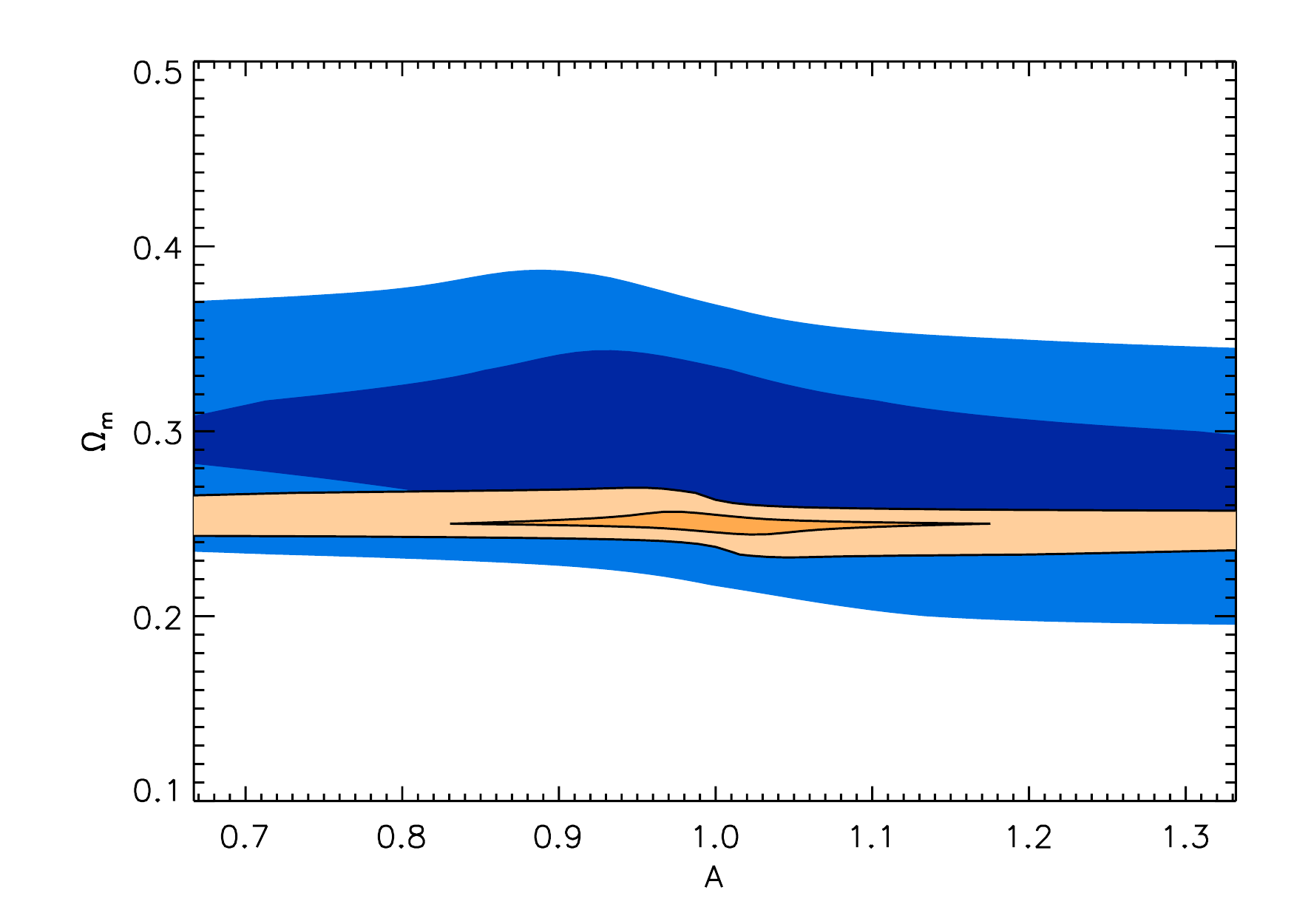}
    \includegraphics[height=1.9in,width=2in]{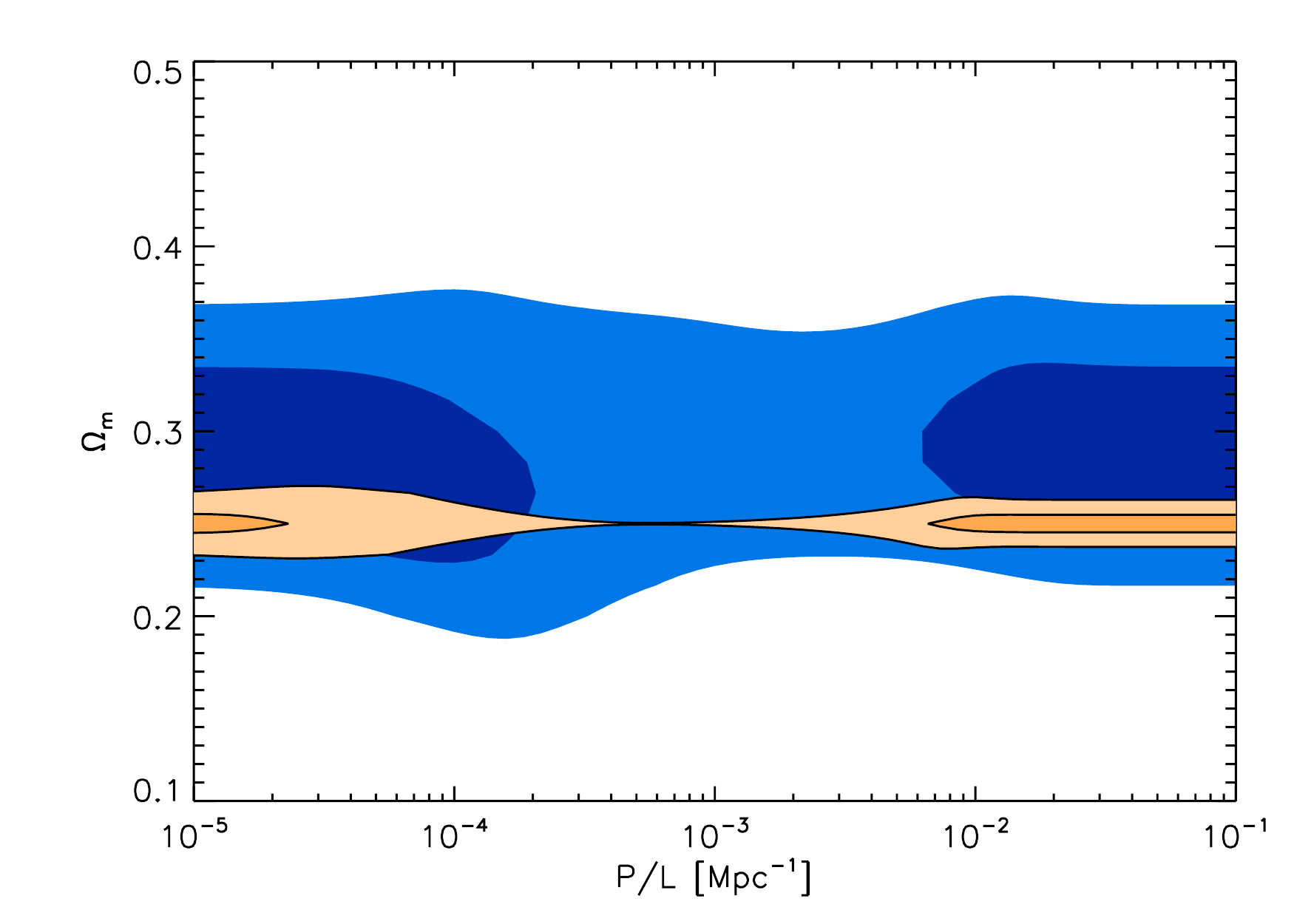}
    \includegraphics[height=1.9in,width=2in]{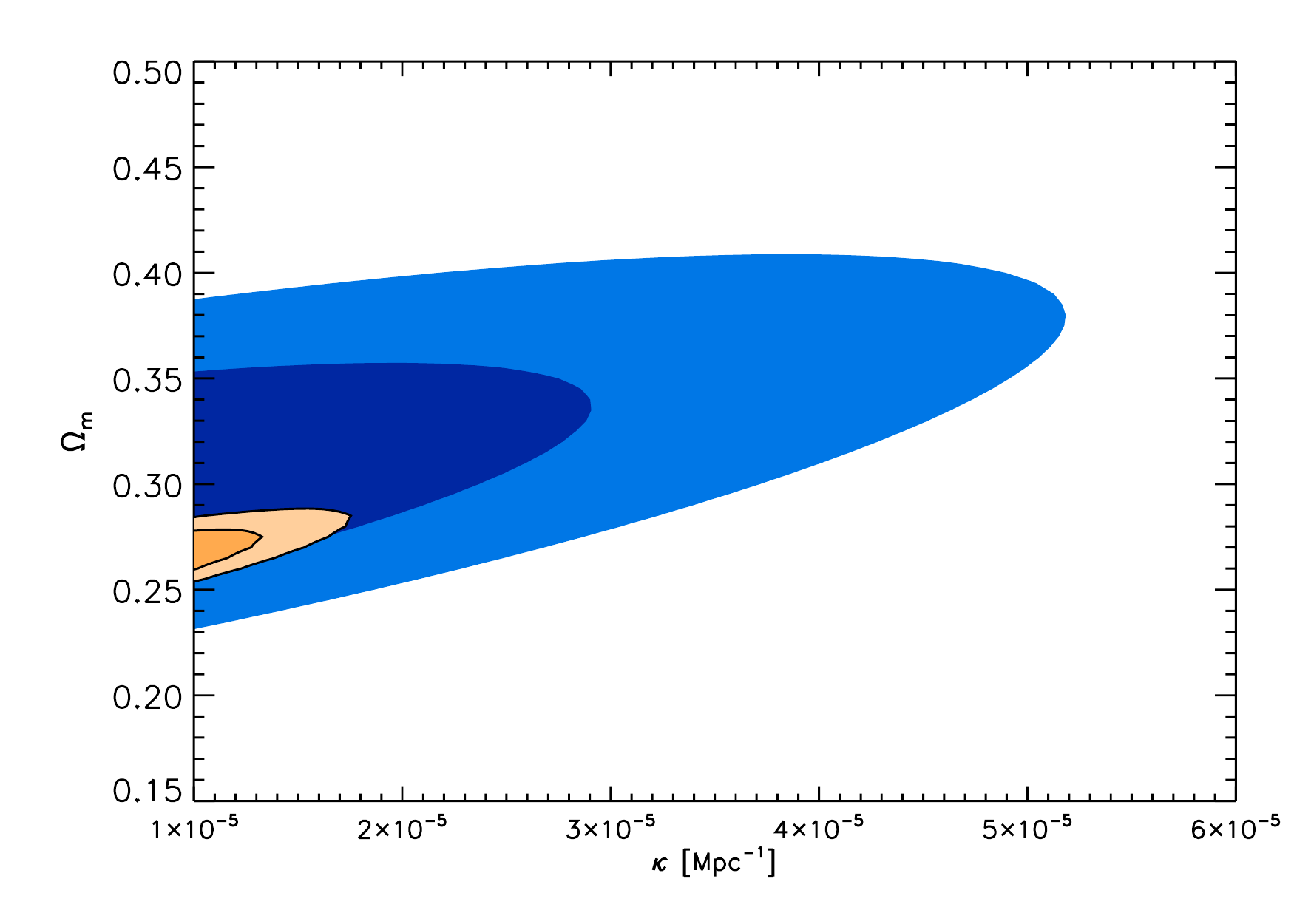} 
    \caption{\label{sumario} Forecast constraints from joint EUCLID 
                   (orange scale), shown together with the corresponding constraints from 
                   current data, namely SN (Union08) joint with chronometer $H(z)$ (blue 
                   scale).  Dark and light contours correspond to 1- and 2-$\sigma$ respectively.  
                   From top left to bottom right: constraints on the opacity parameter $\epsilon$, parameter $P/L$ for the simple ALP model of section, parameters $A$ \& $P/L$ for chameleons, and parameter $\kappa$ for MCPs see Ref.~\cite{alpav} for more details.}
  \end{center}
\end{figure} 

If new particles from physics beyond the standard model couple to
photons then the propagation of light may be altered.  We have reviewed two scenarios for exotic particles  which can
significantly modify the propagation of photons as they pass through
magnetic fields.    Measurements of cosmic opacity are a strong tool to constrain
such scenarios, as interactions between photons and exotic particles
in the magnetic fields of the intergalactic medium leads to a new
source of cosmic opacity.
Uniform deviations from cosmic transparency (i.e. opacity) can be constrained through their 
effects on distance duality, by parameterizing possible deviations from the Etherington relation.
The Etherington relation  implies that, in a cosmology based on a metric theory 
of gravity, distance measures are unique: the luminosity distance is $(1 + z)^2$ times the 
angular diameter distance. Both luminosity distance and angular diameter distance depend on the Hubble parameter $H(z)$, but this relation is valid in any cosmological background where photons 
travel on null geodesics and where, crucially, photon number is conserved. We have   restricted our attention on violations of the Etherington relation arising from the  violation of photon conservation.

More exotic sources of photon-conservation violation involve a coupling of photons 
to particles beyond the standard model of particle physics. We have focused on axion-like particles, new scalar or pseudo scalar
fields which couple to the kinetic terms of photons, and
mini-charged particles which are hidden sector particles with  a tiny electric charge.  Photons
passing through intergalactic magnetic fields may be lost by pair
production of light
mini-charged particles.  If the mixing between axion-like particles
and photons is significant, then interactions in the intergalactic magnetic
fields will also lead to a loss of photons due to conversion into
ALPs.  However if the coupling between photons and ALPs is sufficiently
strong, one-third of any initial flux will be converted into ALPs,
and two-thirds into photons, resulting in a redshift-independent
dimming of supernovae which we cannot constrain or exclude with
cosmic opacity bounds. 

The improved measurement of the cosmic opacity found here leads to
improved bounds on these exotic physics scenarios which are summarised
in Fig. \ref{sumario}.  Future measurements of baryon acoustic
oscillations, and an increase in the number of observations of high
redshift supernovae will lead to further improvements in the
constraints on physics beyond the standard model.

%
%
\end{document}